\documentstyle[prd,aps]{revtex}
\begin{document}
\draft
\twocolumn[\hsize\textwidth\columnwidth\hsize\csname
@twocolumnfalse\endcsname
\preprint{SU-ITP-96/20, SUSSEX-AST 96/5-1,  astro-ph/9605094}
\title{Density Perturbations and Black Hole Formation in Hybrid Inflation}
\author{Juan Garc\'\i a-Bellido$^\ast$, Andrei Linde$^\dagger$ and
David Wands$^{\ddagger\ast}$}
\address{$^\ast$Astronomy Centre, University of Sussex, Falmer,
Brighton BN1 9QH,\ U.K.\\
$^\dagger$Physics Department, Stanford University,
Stanford CA 94305-4060, USA\\
$^\ddagger$School of Mathematical Studies, University of Portsmouth,
Portsmouth PO1 2EG,\ U.K.}
\date{May 15, 1996}
\maketitle
\begin{abstract}
We investigate the recently proposed hybrid inflation models with two
stages of inflation. We show that quantum fluctuations at the time
corresponding to the phase transition between the two inflationary
stages can trigger the formation of a large number of inflating
topological defects. In order to study density perturbations in these
models we develop a new method to calculate density perturbations
in a system of two scalar fields. We show that density perturbations in
hybrid inflation models of the new type can be very large on the scale
corresponding to the phase transition. The
resulting density inhomogeneities lead to a copious production of
black holes. This could be an argument against hybrid inflation models
with two stages of inflation. However, we find a class of models where
this problem can be easily avoided. The number of black holes produced
in these models can be made extremely small, but in general it could be
sufficiently large to have important cosmological and
astrophysical implications.  In particular, for certain values of parameters
these  black holes may constitute the dark matter in the universe. It is also
possible to have
hybrid models with two stages of inflation where the black hole
production is not suppressed, but where the typical masses of the black
holes are very small. Such models lead to a completely different
thermal history of the universe, where post-inflationary reheating
occurs via black hole evaporation.
\end{abstract}
\pacs{PACS numbers: 98.80.Cq \hspace*{2mm} SU-ITP-96-20,
SUSSEX-AST 96/5-1,  RCG-96/07, astro-ph/9605094}

 \vskip2pc]

\section{Introduction}

A period of ``inflation'' or accelerated expansion in the early
universe is an attractive idea in modern cosmology.  Acceleration of
the scale factor could drive the universe towards homogeneity,
isotropy and spatial flatness. However it is the ability of quantum
fluctuations in the fields driving inflation to produce a nearly
scale-invariant spectrum of quantum fluctuations that provides the
most powerful test of the inflationary
paradigm and may allow us to constrain the
physics involved. Cosmological observations allow us to measure the
amplitude and tilt of the primordial density and, possibly,
gravitational wave spectra on scales that would have left the horizon
during inflation.

The first inflationary models such as the old and the new inflationary
universe scenario presumed that inflation began in the false vacuum
state after the high temperature phase transitions in the early
universe \cite{old,new}.  Later it was proposed that all possible
initial conditions should be considered without necessarily assuming
initial thermal equilibrium, and see whether some of these conditions
may lead to inflation.  This scenario was called chaotic inflation
\cite{chaotic}. For many years the idea of chaotic initial conditions
seemed too radical, since it implied a considerable deviation from the
idea of the hot Big Bang. It was argued that for a successful
realization of inflationary theory one should satisfy so-called
``thermal constraints'' \cite{TurStein}. However, gradually it was
understood that the assumption of thermal initial conditions is
neither natural nor helpful for inflationary theory \cite{book}. As a
result, most of the models investigated now belong to the class of
chaotic inflation, which provides the most general framework for
the development of inflationary cosmology.

The simplest models of chaotic inflation include theories with
potentials $V(\phi)$ such as $m^2\phi^2/2$ or
$\lambda\phi^4/4$. Inflation occurs in these theories at $\phi >
M_{\rm P}$. However, one may also consider chaotic inflation near
$\phi = 0$ in models with potentials which could be used for
implementation of the new inflation scenario, such as
$-m^2\phi^2/2+\lambda\phi^4/4$\, \cite{chaotic2}. For brevity, one may
call inflation in such models ``new inflation'', to distinguish it
from inflation at large $\phi$, but strictly speaking these models
also belong to the general class of chaotic inflation models: the
original new inflationary universe scenario based on the theory of
high temperature phase transitions have never been successfully
implemented in realistic theories.

The simplest models of chaotic inflation such as the model
$m^2\phi^2/2$ have many advantages, including natural initial
conditions near the Planck density and the existence of the regime of
eternal self-reproduction of the universe~\cite{book}.
Normalizing the mass scale
by the fluctuations in the microwave background observed by
COBE~\cite{COBE} gives $m\simeq2\times10^{13}$ GeV and the energy
density at the end of inflation is $V(\phi) \simeq (10^{16} \,{\rm
GeV})^4$. At this energy gravitational waves contribute about $10$\%
of the microwave background fluctuations.  The tilt of the density
perturbation spectrum in this model is $n-1\simeq-0.03$.

However, inflation occurs in such models only for $\phi {\
\lower-1.2pt\vbox{\hbox{\rlap{$>$}\lower5pt\vbox{\hbox{$\sim$}}}}\ }
M_{\rm P}$.  It is quite possible to have inflation at $\phi > M_{\rm
P}$ in models with polynomial potentials, but in string theory and
supergravity one often encounters potentials which are extremely steep
at $\phi > M_{\rm P}$. It is not an unsolvable problem, see e.g.
\cite{Goncharov}, but it would be nice to have a simple model where
inflation may occur at $\phi < M_{\rm P}$ as well. It is possible to
achieve this, for instance, in versions of ``new inflation'' with
$V(\phi) \sim -m^2\phi^2/2+\lambda\phi^4/4$.  However, in the simplest
models of such type one has an unacceptably large negative tilt of the
spectrum, unless the amplitude of spontaneous symmetry
breaking is much greater than $M_{\rm P}$ \cite{Lyth}. Thus we return
to the problem of having successful inflation at $\phi < M_{\rm P}$.

There has recently been a lot of interest in the hybrid inflation
scenario \cite{hybrid,LL93,CLLSW,Lazarides,Stewart,SO(10)}. Initial
conditions for inflation in this scenario are not determined by
thermal effects, and thus hybrid inflation also belongs to the general
class of chaotic inflation models. However, hybrid inflation may occur
at values of the scalar fields much smaller than~$M_{\rm P}$. The tilt
of the spectrum in hybrid inflation typically is very small and
positive, giving rise to so-called ``blue spectra''~\cite{MML,Carr}.
The contribution of gravitational waves to the microwave background
anisotropies is usually negligible. The reheating
temperature in this scenario is typically large enough to ensure the
possibility of electroweak baryogenesis, but small enough to avoid the
problem of primordial gravitinos.

It is still a challenge to obtain a natural implementation of this
scenario in the context of supergravity and string theory, but in
globally supersymmetric theories this scenario appears in a very
natural way. A very interesting version  of the hybrid inflation
scenario recently proposed by Randall,  Solja\v ci\'c and Guth
was even called ``supernatural''~\cite{Guth}.

A distinctive feature of hybrid inflation is that it describes the
evolution of two scalar fields, $\phi$ and $\psi$. In the beginning
one of these fields (field $\phi$) moves very slowly, and the second
field may not move at all (though this second condition is not
necessary \cite{Lazarides}). The energy density supporting inflation
is dominated by the false vacuum energy of the field $\psi$. At the
moment when the slowly moving field $\phi$ reaches some critical value
$\phi_c$, it triggers a rapid motion of the field $\psi$, inducing a
transition to a ``waterfall'' regime. Then the energy density of the
field $\psi$ rapidly decreases, and inflation ends.

Care is needed in evaluating the spectrum of density perturbations
produced by inflation in the presence of more than one field. Many of
the usual simplifying assumptions break down. Perturbations may no
longer be purely adiabatic and hence curvature perturbations
depend not only on the field fluctuations at horizon crossing but
also upon their subsequent evolution up to the end of inflation, or
even beyond.  In the first versions of hybrid inflation   the
mass of the field $\psi$ whose false vacuum energy density drives
inflation was much larger than that of the slowly-rolling field
$\phi$ and so the single-field approximation was quite sufficient.
Also, inflation ended abruptly when the field $\phi$ reached its
critical value $\phi_c$, and the field $\psi$ began its motion.

This regime is certainly not the most general. Recently
attention has been drawn to the possibility that both fields $\phi$
and $\psi$ could have masses close to the SUSY breaking scale
$m\simeq1$ TeV, while the symmetry breaking scale in the $\psi$
direction may be as large as $M_{\rm P}$ \cite{Guth}. In this case the
process of rolling of the field $\psi$ towards its minimum may take a
lot of time even if its mass is a few times greater than the Hubble
constant at the end of inflation. This may be attractive in the
context of supersymmetric theories but raises new issues about the
generation of density perturbations in the two-field model. We will
consider these issues in this paper and spell out the dangers of
ending hybrid inflation by a slow phase transition.

The main problem associated with this scenario can be explained in
the following way.  The effective potential of the field $\psi$ used
in \cite{Guth} is symmetric with respect to the change $\psi \to
-\psi$. As a result, at the moment of the phase transition the field
$\psi$ can roll with equal probability towards its positive and
negative values. This leads to the usual domain wall problem. To avoid this
problem in the original hybrid inflation model \cite{hybrid} it was
suggested to change the topology of the vacuum manifold and couple
the field $\psi$ to gauge fields. In such a case instead of domain
walls one may obtain either strings or monopoles, or (as in the
electroweak theory) no stable topological defects at all. Monopoles
should be avoided, but strings  do not lead to any cosmological
problems in theories with a relatively small scale of symmetry
breaking, as studied in Ref.~\cite{hybrid}.
Alternatively, one may consider versions of the hybrid inflation
scenario considered in Ref.~\cite{Lazarides}, where no topological
defects are produced.

In the model proposed in Ref.~\cite{Guth} topological defects do
appear.  In the simplest realization of this model one gets domain
walls, which should be avoided at all costs. If one modifies the model
to produce strings instead, one also has a problem, since strings
corresponding to the scale of spontaneous symmetry breaking $\sim
M_{\rm P}$ by themselves produce density perturbations
$\delta\rho/\rho \sim 1$ on all scales.  One could expect that
monopoles would not lead to any trouble since the distance between
them grows exponentially during the second stage of
inflation. However, because of inflation, which occurs in this model
during the long stage of rolling of the field $\psi$ to its minimum,
all topological defects in this model appear to be inflating, as in
\cite{LL}. Independently of the nature of these defects (domain walls,
strings, monopoles, either topologically stable or not) their
exponential expansion leads to density perturbations $\delta\rho/\rho
= {\cal O}(1)$ on the exponentially large scale corresponding to the
moment of the phase transition. This may result in a copious black
hole formation. However, inflating topological defects in our model
are rather specific, because they appear in the theory with $|m_\psi|
> H$. For this reason the possibility of black hole formation by such
defects requires separate investigation. This problem is extremely
interesting since here for the first time the issue of inflating
topological defects appears in the context of observational cosmology.

Independently of this issue,
the appearance of inflating topological defects clearly demonstrates that the
existence of the second stage of inflation in the hybrid scenario may lead to
large density perturbations. As pointed out in Ref.~\cite{Guth}, the phase
transition at $\phi = \phi_c$ leads to the appearance of a characteristic
spike in the spectrum of density perturbations. The existence of such a
spike was first found in a similar model by Kofman and
Pogosyan~\cite{Kofman}. In the ``supernatural'' hybrid inflation model
it is difficult to calculate the amplitude of the peak of the
spectrum; in Ref.~\cite{Guth} it was done slightly away from the point of the
phase transition, where the amplitude of the density perturbations
has already diminished. In order to perform the calculation, the evolution of
the fluctuating field in \cite{Guth} was divided into several parts, and
different approximations were used at every new step. However, the results of
calculations of the amplitude of density perturbations near the narrow peak can
be very sensitive to the choice of the approximation, especially in a situation
where one may expect density perturbations to be large.  Therefore  we
developed a more direct method of calculations of density perturbations in this
model.

One of the most interesting aspects of the model of Ref.~\cite{Guth}
is the existence of a regime in which quantum diffusion of the
coarse-grained background fields dominates over its classical
evolution and determines prominent features within our present
cosmological horizon. Previously such phenomena were confined to
scales much beyond our present horizon and were usually ignored.
In Ref.~\cite{Guth}, the machinery of stochastic inflation, see
Ref.~\cite{Stochastic}, was used to estimate the behavior of the
fields close to the phase transition, where large quantum fluctuations
make the stochastic formalism necessary. In this paper we will use
this formalism to find whether or not most of inflationary trajectories come
through the region where large density perturbations are generated.  We believe
that the method  which we developed may be of interest in
its own right and can be applied to a  more general class of models with many
scalar fields.

Our final
results  agree with the conclusion based on
the topological defect analysis:   density perturbations created at the
moment of the phase transition are very large. In particular, in the model of
Ref.~\cite{Guth} with the parameters given there corresponding to the
second stage of inflation lasting for 20 - 30 Hubble times $H^{-1}$, one has
density perturbations $\delta\rho/\rho = {\cal O}(1)$. In such a situation one
can expect copious production of huge black holes, which should lead to
disastrous cosmological consequences.

However, this is not an unsolvable problem. For a suitable choice of
parameters the second stage of inflation can be completely eliminated,
and in this respect the model can be made very similar to the original
hybrid inflation model of Ref.~\cite{hybrid}, where the problem of
black holes does not appear at all. A very interesting possibility
appears if the second stage of inflation does exist, but is very
short, lasting only two or three Hubble times $H^{-1}$. Then the black
holes formed from the large density perturbations may be small enough
to evaporate quickly. With the parameters of the model of
Ref.~\cite{Guth} the evaporation time is still very large even for the
smallest black holes. However, if one studies hybrid inflation with a
larger Hubble constant, the black holes produced during inflation can
be made very small, so that they evaporate before nucleosynthesis.
Even if the probability of formation of such black
holes is suppressed, the fraction of matter in such micro black holes
at the moment of their evaporation may be quite substantial, since the
fraction of energy in radiation rapidly decreases in an expanding
universe.  This may lead to crucial modifications of the thermal
history of the universe and may rejuvenate the possibility that the
baryon asymmetry of the universe was produced in the process of black
hole evaporation~\cite{Hawk,Liddle}.

Finally, one may consider the models where the second stage of inflation
lasts for about ten Hubble times. As we will see, in this case the probability
of the black hole formation may be sufficiently small, so that the amount of
black holes does not contradict the cosmological bounds on the black hole
abundance. It raises a very interesting possibility (see also \cite{NovNas})
that the black holes
produced in the hybrid inflation scenario may serve as the dark matter
candidates. In other words, dark matter may indeed be black!

As we already mentioned, there exist some versions of the hybrid
inflation scenario where topological defects do not appear at all. In
this paper we will suggest another version of such a scenario, which
we will call ``natural'' hybrid inflation. This scenario is a hybrid
of the simplest version of ``natural inflation'' \cite{BG,Natural}, and
the model of Ref.~\cite{Guth}.  It shares some attractive features
of ``natural inflation'' such as the natural origin of small
parameters appearing in the theory. On the other hand, unlike the
original ``natural inflation'', our scenario does not require the
radius of the ``Mexican hat'' potential to be greater than the Planck
scale, which causes problems when one attempts to implement ``natural
inflation'' in string theory \cite{BG}. We will show that in the
models of natural hybrid inflation one can easily avoid the problem of large
density perturbations.

The plan of the paper is as follows. In Sect. 2 we will briefly
describe the simplest hybrid inflation model \cite{hybrid} and its
relation to the model of Ref.~\cite{Guth}. We will find classical
solutions describing the evolution of the fields $\phi$ and $\psi$ in this
model. Most of our investigation will be fairly general, but since the
original hybrid inflation model \cite{hybrid} is already well
investigated~\cite{CLLSW}, we will concentrate on the model of
Ref.~\cite{Guth} where an additional stage of inflation occurs after
the phase transition. In Sect. 3 we will evaluate the amplitude of
quantum fluctuations of each scalar field. In Sect. 4 we analyze the
issue of inflating topological defects and the associated
density perturbations. In Sect. 5 we study density perturbations both
before and after the phase transition and then discuss the important
issue of quantum diffusion at the phase transition. In Sect. 6 we
analyze the probability of primordial black hole formation due to
large density perturbations. We will also discuss the possibility of
reheating of the universe by evaporation of small black holes. In
Sect. 7 we propose and briefly describe the ``natural'' hybrid
inflation model. We will discuss our results and summarize our
conclusions in Section 8.

\section{Classical field dynamics}
\label{DYN}

The simplest realization of chaotic hybrid inflation is provided by
the potential~\cite{hybrid}
\begin{equation}\label{hybrid}
V(\phi,\psi) = \left(M^2-{\sqrt{\lambda}\over2}\psi^2 \right)^2
 + {1\over2}m^2\phi^2 + {1\over2}\gamma\phi^2\psi^2 \, .
\end{equation}
For comparison, we will write here the effective potential of one of
the models considered in Ref.~\cite{Guth}:
\begin{equation}\label{supernatural}
V(\phi,\psi) =
M^4 \cos^2\left({\psi\over\sqrt2 f}\right)
+ {1\over2}m^2\phi^2
+ {1\over4}\lambda^2 \phi^2  \psi^2\, .
\end{equation}
In the region $\psi < f$, where inflation occurs in the model
(\ref{supernatural}), the potentials (\ref{hybrid}) and
(\ref{supernatural}) practically coincide, with the redefinition of
parameters, $f^2 \to M^2/2\sqrt\lambda$ and $\lambda^2 \to 2\gamma$.
(Note that the fields denoted $\phi$ and $\psi$ in Ref.~\cite{Guth}
correspond to our fields $\psi$ and $\phi$.  In this respect we have
followed the notation of Ref.~\cite{CLLSW}.) In what follows we will study
the model (\ref{hybrid}), but we will be most interested in values
of the parameters close to those of Ref.~\cite{Guth}.

The equations of motion for the homogeneous fields are then
\begin{eqnarray}\label{EQM}
\ddot\phi + 3H\dot\phi
 &=& - (m^2 + \gamma\psi^2) \phi\, , \\
\ddot\psi + 3H\dot\psi
 &=& (2\sqrt\lambda M^2 - \gamma\phi^2 - \lambda\psi^2) \psi \, ,
\end{eqnarray}
subject to the Friedmann constraint,
\begin{equation}
H^2 = {8\pi\over3M^2_{\rm P}} \left[ V(\phi,\psi)
 + {1\over2}\dot\phi^2 + {1\over2}\dot\psi^2 \right] \, .
\end{equation}
Although we can always integrate the equations of motion numerically
for arbitrary initial conditions, the classical motion of the
homogeneous field is not necessarily a good representation of the
coarse-grained field on super-horizon scales. If the classical motion
is sufficiently slow it can become dominated by quantum diffusion
caused by wave-modes crossing outside the Hubble radius.

In particular we wish to consider the case when $\phi$ is much larger
than $\psi$ at early times so that $\psi$ has a large positive mass
$\simeq\gamma\phi^2$ and rolls rapidly to $\psi=0$.\footnote{If
instead $\phi$ rolls rapidly to $\phi=0$ it remains there and the
problem reduces to the usual case where the $\psi$ field rolls
directly to the global minimum at $V=0$.} For values of $\phi$ above
a critical value $\phi_c$, $\psi=0$ is a stable minimum. Thus $\psi$
remains zero while $\phi$ slowly-rolls (for $m^2\ll H^2$) down to the
critical value, $\phi=\phi_c$, where the $\psi$ field becomes
massless. For smaller values, $\psi=0$ is an unstable local maximum
and quantum diffusion initiates a second-order phase transition from
the false vacuum to the true vacuum state.

In the simplest version of hybrid inflation, where the couplings
$\lambda$ and $\gamma$ are of order unity, this is essentially all the
dynamical evolution that matters. The bare mass of the $\psi$ field,
$-m_\psi^2=2\sqrt\lambda M^2$, must be much larger than
$H^2$~\cite{CLLSW} and the phase transition occurs rapidly and
inflation ends. The perturbation constraints on large scales are then
readily derived from the usual single field results where the role of
$\psi$ at early times can be neglected.

But what if the bare mass of the $\psi$ field is not very much larger
than $H$? In particular, what if this field has the type of potential
we might expect for a moduli field with a minimum at $\psi\sim
M_{\rm P}$ and a negative mass-squared $-m_\psi^2$ of order
$(1 {\rm TeV})^2$ about $\psi=0$?
The false vacuum energy density at $\psi=0$ is then
$M^4\sim m^2M_{\rm P}^2\simeq(10^{11} {\rm GeV})^4$ and the Hubble
constant $H\simeq m_\psi\sim1$TeV, as discussed in Ref.~\cite{Guth}.
It lies outside the range of parameters originally considered for
hybrid inflation since it corresponds to an exceedingly flat potential
for $\psi$ with effective coupling constant $\lambda\sim
m_\psi^4/M^4\sim10^{-30}$.  The
roll-down of the $\psi$ field need no longer be fast and we must
consider the complicated evolution of both the fields $\phi$ and
$\psi$ during this stage.

When $\psi=0$ the potential simply reduces to
$V(\phi)=M^4+m^2\phi^2/2$. For the range of parameters and field
values that we are interested in the constant term always dominates
and the Hubble expansion can be taken to be de Sitter expansion with
$H=H_0\equiv\sqrt{8\pi/3}\,M^2/M_{\rm P}$. It is useful to then write
the bare masses of the two fields $\phi$ and $\psi$ relative to the
Hubble scale as
\begin{equation}
\alpha \equiv {m^2\over H_0^2} \hspace{8mm} {\rm and} \hspace{8mm}
\beta \equiv 2\sqrt\lambda \, {M^2\over H_0^2} \, .
\end{equation}

In the case of a single scalar field $\sigma$ evolving during
inflation, one usually resorts to the slow-roll, $\dot\sigma^2\ll
V(\sigma)$, and quasi-massless, $V''\ll H^2$, approximations to make
analytic progress. This allows one to reduce the equations of motion
for the scalar field to a first-order equation. However in our case
the mass of the $\psi$ field is less than $H$ only for a short interval,
$\phi_c\sqrt{1-1/\beta}<\phi<\phi_c\sqrt{1+1/\beta}$. Even then, we
wish to consider values of $\alpha$ not much below unity so the
quasi-massless approximation may not be very good for $\phi$ either.

Fortunately the fact that the potential energy, and hence the Hubble
rate, are so nearly constant (and this really is a {\rm very} good
approximation for the parameters of Ref~\cite{Guth}) allows us to
integrate the second order-equations in two regimes.

\subsection*{Region~I: small $\psi$}

The first approximation regime will be for
$\,\psi^2/H_0^2\ll\beta/\gamma$.\footnote
{We also require $\psi^2/H_0^2\ll\beta/\lambda$ but in
practice this is a much weaker condition for the parameters
of~\cite{Guth}.} This leaves the mass of the $\phi$ field
constant and the equation of motion becomes
\begin{equation}
\phi'' + 3\phi' + \alpha \phi = 0 \, ,
\end{equation}
where a prime denotes a derivative with respect to $N=H_0(t-t_e)$,
the number of $e$-folds to the end of inflation. This can
be readily integrated to give
\begin{eqnarray}\label{r}
\phi(N) & = & \phi_+ \exp(-r_+N) + \phi_- \exp(-r_-N) \,,\nonumber\\
r_\pm & = & {3\over2} \mp \sqrt{{9\over4}-\alpha}\,.
\end{eqnarray}
The asymptotic solution is $\phi = \phi_+ \exp (-r N)$ where
$r\equiv r_+>0$ which approaches the slow-roll solution $\phi = \phi_+
\exp (-\alpha N/3)$ for $\alpha\ll1$.

For $\phi>\phi_c$, the $\psi$ field remains
trapped in the stable minimum at $\psi=0$ and we have effectively
single-field inflation and the above solution gives the
exact evolution of $\phi(N)$.

Below $\phi\approx\phi_c$ we can no longer take the $\psi$ field to
remain fixed at $\psi=0$. The $\psi$ field equation of motion is
\begin{equation}
\psi'' +3\psi'
 = \left(\beta - \gamma{\phi^2\over H_0^2} \right) \psi \, .
\end{equation}
We see that the field becomes massless at
$\phi_c^2\equiv(\beta/\gamma)H_0^2$.
Re-writing this as an equation for $\psi$ as a function of $\phi$ gives
\begin{equation}
\phi^2 {d^2\psi\over d\phi^2} + (1-2q)\phi{d\psi\over d\phi}
+ (\kappa^2\phi^2 + q^2 - \nu^2)\psi = 0 \, .
\end{equation}
where
\begin{equation}\label{nu}
q \equiv {3\over2r} \, , \hspace{5mm}
\kappa \equiv {\sqrt{\beta}\over r}\, {1\over\phi_c} \, , \hspace{5mm}
\nu \equiv {1\over r} \sqrt{{9\over4}+\beta}\,.
\end{equation}
The exact solution is a linear combination of Bessel functions,
\begin{equation}
\psi(\phi) = \phi^q \left[ - c_1 Y_\nu(\kappa\phi) + c_2 J_\nu(\kappa\phi)
\right] \, .
\end{equation}
For $\phi\ll\phi_c$ the growing mode is given by the small angle
approximation
\begin{eqnarray}
\label{growpsi}
\psi(\phi) &=& - c_1 \,\phi^q \,Y_\nu(\kappa\phi) \approx
c_1 \,A\, \phi^{-(\nu-q)}\\ \label{psiN1}
&=& c_1 \,A\, \phi_+^{-(\nu-q)}\, e^{(\nu-q) r\,N}\,,
\end{eqnarray}
where the numerical coefficient $A=(2/\kappa)^\nu\Gamma(\nu)/\pi$.

\subsection*{Region~II: small $\phi$}

As $\phi$ decreases, the effective potential for the $\psi$ field
becomes dominated by its bare (tachyonic) mass $\sqrt{\beta}H_0$.
Thus for $\phi\ll\phi_c$, and still assuming
$\,\psi^2/H_0^2\ll \beta/\lambda$, we have
\begin{equation}
\psi'' + 3\psi' - \beta \psi = 0\,,
\end{equation}
which has the general solution
\begin{eqnarray}
\psi(N) & = & \psi_+ \exp(s_+ N) + \psi_- \exp(s_- N)\,,\nonumber\\
\label{pm}
s_\pm & = & -{3\over2} \pm \sqrt{{9\over4} + \beta}\,.
\end{eqnarray}

Matching to the asymptotic solution in Region~I, Eq.~(\ref{psiN1}),
we see that only the growing mode ($s=s_+ > 0$) exists in Region~II.
Recalling that $N$ is measured from the end of inflation, it is
simply given by
\begin{equation}
\label{psiN}
\psi(N) = \psi_e \exp(sN) \, ,
\end{equation}
for all trajectories. The total amount of inflation after a given
point is determined solely by the ratio $\psi_e/\psi$, as noted in
Ref.~\cite{Guth}.
Even though the effective mass of $\psi$ has become large and
negative, inflation will only end when $\dot\psi^2\simeq V$, which
implies\footnote
{Note that for $\beta>1$ this also ensures that de Sitter remains a
good approximation until very near the end of inflation.}
\begin{equation}
\psi^2 \simeq \psi_e^2 \equiv \, {M_{\rm P}^2\over4\pi s(s+1)}  \, .
\end{equation}
{}From a given value $\psi$, the evolution will take
\begin{equation}
\label{Npsi}
N(\psi) \approx {1\over s} \ln\left({M_{\rm P}\over s\,\psi}\right)
\end{equation}
$e$-folds to the end of inflation.  As we shall see, if the initial
value of $\psi$ is only of order $H_0\sim1$TeV
there must be a further $(32/s)$ $e$-folds before inflation ends.

The growing value of $\psi$ increases the effective mass of the
$\phi$ field through the interaction term and it also becomes larger
than $H$. However, the $\phi$ field is much closer to the minimum of
its potential than the $\psi$ field ($\phi_c/M_{\rm P}\sim10^{-12}$)
and $\phi$ soon starts to execute damped oscillations about
$\phi=0$, as noted in Ref.~\cite{Guth}.

This time we have a parametric equation for the inflaton field in terms
of the triggering field, for which the general solution is
\begin{equation}\label{exactphi}
\phi(\psi) = \psi^p\left[\bar{c}_1 J_\mu(\rho \psi) +
\bar{c}_2 Y_\mu(\rho \psi)\right]\,,
\end{equation}
where
\begin{equation}\label{param}
p \equiv -{3\over2s}\,,\hspace{5mm}
\rho \equiv {\sqrt\beta\over s}\, {1\over\phi_c}\,,\hspace{5mm}
\mu \equiv {1\over s} \sqrt{{9\over4} - \alpha}\,.
\end{equation}

Matching to the asymptotic solution in Eq.~(\ref{growpsi}), we find
that only the $J_\mu$ solution is selected. We can see this from
the small $\rho\psi$ expansion of Eq.~(\ref{exactphi}) with
$\bar{c}_2=0$,
\begin{equation}
\phi(\psi) = \bar{c}_1\, \psi^p\, J_\mu(\rho \psi)  \simeq
\bar{c}_1\, B \,\psi^{p+\mu}\,,
\end{equation}
where $B=(\rho/2)^\mu/\Gamma(\mu+1)$ and
$p+\mu \equiv -(\nu-q)^{-1} \equiv -\,r/ s$.
Comparing this with the limiting behavior of Eq.~(\ref{growpsi})
we find a relation between the coefficients
\begin{equation}
\bar{c}_1\,B = (c_1\, A)^{s/r}\,.
\end{equation}
Therefore, as long as the solutions pass through the overlapping region,
our solutions of Region~I evolve smoothly into Region~II.

\section{Quantum field fluctuations}

In this section we discuss the evolution of quantum fluctuations
of the fields. In the slow-roll approximation, the amplitude of
quantum fluctuations of a massless field at horizon crossing ($k=aH$)
is approximately $H/2\pi$. However, in our case the masses of the
fields are not necessarily much smaller than the Hubble scale, and
corrections to the slow-roll result could be large.

Since the potential $V(\phi,\psi)\simeq V_0$ to a very good
approximation, we can neglect the gravitational backreaction of the
fields, and the equations of motion for linear perturbations in
$\psi$ and $\phi$ can be written as
\begin{eqnarray}
\ddot{\delta\psi} + 3H \dot{\delta\psi} &+& \left({k^2\over a^2} -
\beta H^2 + \gamma\phi^2\right)\delta\psi \nonumber\\
&=& - 2 \gamma\psi\phi\delta\phi  - 3\lambda\psi^2\delta\psi\,,\\
\ddot{\delta\phi} + 3H \dot{\delta\phi} &+& \left({k^2\over a^2} +
\alpha H^2\right)\delta\phi \nonumber\\
&=& - \gamma\psi^2\delta\phi - 2 \gamma\psi\phi
\delta\psi \,.
\end{eqnarray}
Note that in Region~I, when $\psi=0$, the terms in the
right-hand-sides are zero and the evolution of $\delta\psi$ and
$\delta\phi$ decouple. We can write these equations, in terms of
the canonically quantized fields
$u\equiv a\delta\psi$ and $v\equiv a\delta\phi$, as
\begin{eqnarray}\label{uv}
u_k'' + \left(k^2 - {2 + \beta(1-\eta^{2r})\over\eta^2}\right)
u_k = 0\,,\\
v_k'' + \left(k^2 - {2 - \alpha\over\eta^2}\right) v_k = 0\,,
\end{eqnarray}
where primes denote derivatives with respect to conformal time,
$\eta=-1/aH$, and we have chosen $\eta=-1$ when $\phi=\phi_c$.

Since the mass of the $\phi$ field is constant, we can write an exact
expression for the quantum fluctuations,
\begin{equation}
v_k(\eta) = {\sqrt\pi\over2\sqrt k}\,e^{i(1-r)\pi/2}\,
(-k\eta)^{1/2}\,H_{3/2-r}^{(1)}(-k\eta)\,,
\end{equation}
where $r=r_+$ is defined in Eq.~(\ref{r}). This has the correct
flat-space limit as $-k\eta\to\infty$, $v_k\to e^{ik\eta}/\sqrt{2k}$,
while as $\phi\to0$, and $-k\eta\to0$, we find
\begin{equation}
v_k(\eta) = {C(r)\over\sqrt{2k}}\,e^{i(1-r)\pi/2}\,
(-k\eta)^{r-1}\,,
\end{equation}
where~\cite{SL}
\begin{equation}\label{Cr}
C(r) = 2^{-r}\,{\Gamma(3/2-r)\over\Gamma(3/2)}\,.
\end{equation}
This results in a scale-invariant spectrum of the growing-mode
perturbations at horizon crossing with amplitude
\begin{equation}\label{deltaphi}
\delta\phi_* = C(r)\,{H\over2\pi}\,.
\end{equation}
Note that the coefficient $C(r)$ gives a constant correction
(independent of scale) to the usual amplitude of curvature perturbations
(obtained in the slow-roll limit where $C(0)=1$).

On the other hand, the effective mass of $\psi$ changes with time and
we cannot write down an exact solution. However we can understand the
qualitative behavior by considering the effective Schr\"odinger
equation for $u_k$, see Eq.~(\ref{uv}), with the time-dependent potential
$V(\eta) = - (2+\beta)/\eta^2 + \beta/\eta^{2(1-r)}$.
This has a maximum value $V_{\rm max}$ at $\eta_{\rm max}$ given by
\begin{eqnarray}
V_{\rm max} &=&
 {(\beta+2)r\over1-r}\,\left[{\beta(1-r)\over\beta+2}\right]^{1/r} \,,\\
\eta_{\rm max} &=& \left[{\beta+2\over\beta(1-r)}\right]^{1/2r} \,.
\end{eqnarray}
On small scales, $k^2\gg V_{\rm max}$, the field $u_k$ oscillates
with almost constant amplitude until $-\eta\lesssim\sqrt{\beta+2}/k$,
when it starts to diverge. As $k\to\infty$ we recover the
constant-mass asymptotic ($\eta\to0$) solution
\begin{equation}
u_k(\eta) = {C(-s)\over\sqrt{2k}}\,e^{i(1+s)\pi/2}\,
(-k\eta)^{-s-1}\,,
\end{equation}
where $s=s_+$ is defined in Eq.~(\ref{pm}). However, the amplitude
of the quantum fluctuations decay exponentially for
$k^2<V(\eta)$. Thus modes with $k^2<V_{\rm max}$ will be suppressed.
The amplitude of the growing mode of the field perturbations at
horizon crossing can then be written as
\begin{equation}\label{Ck}
\delta\psi_* = C_k \,{H\over2\pi}\,.
\end{equation}
where the coefficient $C_k$ is scale-dependent, as shown in Fig.~1
for the case $\beta=8$.

We can understand qualitatively the behavior of this
growing amplitude of quantum fluctuations. For modes $k\simeq1$
that leave the horizon near $\phi\simeq\phi_c$, the $\psi$ field
is effectively massless and the amplitude of quantum fluctuations
has the usual value $H/2\pi$, with coefficient $C_k\simeq1$.
When $\phi\lesssim\phi_c\,(1-1/\beta)^{1/2}$, corresponding to
scales $k>1+1/2r\beta$, the magnitude of the (imaginary) mass of
the $\psi$ field becomes larger than $H$.
Then the amplitude of quantum fluctuations even before horizon
crossing, when $k/a\simeq m_\psi$, is $\delta\psi\sim m_\psi/2\pi$,
which is already greater than $H/2\pi$.
As $\phi$ decreases and the corresponding scale increases, the effective
mass of the $\psi$ field approaches its bare value while the amplitude
of quantum fluctuations grows. At very large $k$ we recover the
constant-mass scale-invariant value, $C_\infty = C(-s)$, see
Eq.~(\ref{Cr}). For $\beta=8$, this asymptotic value is
$C_\infty=8.91\gg1$, see Fig.~1. On the other hand, for $k^2<V_{\rm
max}$, the mass (squared) of the $\psi$ field is large and positive at
horizon crossing and the amplitude of these quantum fluctuations is
suppressed.

\section{Inflating Topological Defects}\label{defects}

During inflation we must also consider the effect of short wavelength
fluctuations that cross outside the horizon and perturb the
coarse-grained background field on super-horizon scales.
One can get a pretty good idea of the amplitude of
perturbations on a scale corresponding to the time of the phase
transition by investigating the inflating topological defects produced
at that time.  For this purpose
one should study the evolution of the fluctuations $\delta\psi$ at
the moment of the phase transition.

Before the phase transition the field $\psi$ is very heavy and its
quantum fluctuations can be neglected. As we have seen, its
fluctuations are generated when the mass of the $\psi$ field
becomes smaller than $H$, i.e. close to the phase transition.
The exact duration of this stage is strongly
model-dependent, but with the parameters used in Ref.~\cite{Guth}
the time before the phase transition when the fluctuation can grow is
only about $H^{-1}$. As a result, one may (approximately) visualize
the scalar field $\psi$ at the moment of the phase transition as a
sinusoidal wave with wavelength $H^{-1}$ and amplitude
$\delta\psi \sim H/\sqrt 2\pi$,
\begin{equation}\label{top1}
\delta\psi_c(x) = { H\over\sqrt 2\pi }\, \sin Hx \, .
\end{equation}
(Note that the amplitude here is $H/\sqrt 2\pi$, whereas the
standard expression $H/2\pi$ holds for the averaged amplitude
$\langle\delta\phi^2\rangle^{1/2}$, i.e. for the dispersion of the field.)

This representation is not exact since many different waves give
a contribution to $\delta\psi(x)$, but their wavelengths are
comparable and, for the models we consider, the amplitude of
$\delta\psi$ in Eq.~(\ref{top1}) is indeed of the order of $H/\sqrt2\pi$.
During inflation the wavelength of these perturbations
grows exponentially, $\lambda \sim H^{-1}\, a(t)/a(t_c)$,
where $a(t_c)$ is the scale factor at the moment of the phase
transition, but the amplitude of the field $\delta\phi$ also grows
exponentially.  This is the main reason why spontaneous symmetry
breaking occurs despite the fact that formally the field $\psi$,
averaged over the whole universe, always remains equal to zero.
During each interval of time $\Delta t$, when the effective potential
can be represented as $V_0-m^2(t)\psi^2/2 $ with $m^2(t) \ll H^2$,
the amplitude of the field $\delta\psi$ in (\ref{top1}) grows as
\begin{equation}\label{top2}
\delta\psi_c \sim { H\over \sqrt  2\pi }\exp{m^2  \Delta t \over 3 H}\ \sin
\left({a(t)H\,x \over a(t+\Delta t)}\right) \ .
\end{equation}
whereas for $m^2(t) \gg H^2$ one has
\begin{equation}\label{top3}
\delta\psi_c \sim { H\over \sqrt 2\pi}\exp{(m\, \Delta t) }\ \sin
\left({a(t)H\,x\over a(t+\Delta t)}\right) \ .
\end{equation}
In both cases spontaneous symmetry breaking during inflation
preserves the simple sinusoidal shape of the perturbation, until the
field rolls down to its minimum and a more complicated nonlinear
regime begins. But this occurs already at the very end of inflation.

Note that inflation near the cores of the topological defects continues for a
while even after the field $psi$ reaches its minimum at $\psi = f \sim M_{\rm
P}$. Indeed,   at that time the gradient energy density of the field
distribution (\ref{top3}) ia $\sim H^2 M_{\rm P}^2 \exp(- 2N_c)$, where the
factor $\exp(- 2N_c)$ appears because of the stretching of the wave during
inflation. For large $N_c$ it is much smaller than  the potential energy
density of the field $\psi$, given by
$V(0) = 3H^2 M_{\rm P}^2/8\pi$. As for   the kinetic energy density of the
field $\psi$ it  is always small near $\psi = 0$. Therefore inflation in the
vicinity of of topological defects continues for a long time, until eventually
the gradient energy becomes greater than $V(0)$.

As a result, after the field (\ref{top3}) approaches the minimum of the
effective potential, on the exponentially large scale $\lambda \sim H^{-1}\,
e^{N_c}$ we will have an extremely inhomogeneous matter
distribution. Roughly speaking, in half of the volume of the
universe on that scale, the field will be near the minimum of its
effective potential, whereas in the other half the scalar field
will be close to the top of the effective potential with $\psi = 0$,
and inflation will be still going on. This shows that we will
have density perturbations $\delta\rho/\rho = {\cal O}(1)$ on
exponentially large scales corresponding to the time of the phase
transition.

This effect is very general. It is related to the inflating topological
defects discovered in Ref.~\cite{LL}. The field may roll in any direction
from $\psi = 0$, but stable regions with $\psi = 0$ are constantly
being created, corresponding to inflating domain walls, strings, or
monopoles. However, in Ref.~\cite{LL} the curvature of the effective
potential was much smaller than $H^2$. In our case the second stage
of inflation occurs even if $m_\psi^2>H^2$, because the initial value
of the field $\delta\psi_c(x) =  H/\sqrt 2\pi$ is much smaller than the
amplitude of spontaneous symmetry breaking, so it takes a lot of time
for the field $\psi$ to roll down. This specific feature implies that
there will be no eternal self-reproduction and no fractal structure of
topological defects in our model.

Note that a similar effect may occur even in the models where
topological defects are unstable if the decay rate of unstable defects
is sufficiently small~\cite{Lepora}.  A possible example may be
provided by the metastable electroweak cosmic strings or other
``embedded defects'' \cite{Vachaspati}. Such defects are unstable and
usually do not cause any cosmological problems. However, they also
have $\psi = 0$ in their cores. If they inflate~\cite{Lepora} and
decay only after the end of inflation, they lead to large density
perturbations in the same way as the topologically stable inflating
defects.

One may expect that large density inhomogeneities should lead to a
copious black hole formation. The problem of black hole formation in
a hot universe is not trivial because pressure gradients do not
allow black holes to be formed unless density perturbations on the
horizon scale are of the order of one or even
greater~\cite{HCarr,Khlopov,Carr,Guth,Novikov,NovFrol}. But this is
precisely the case for inflating topological defects. The centers of
these defects remain forever at large density near the top of the
effective potential, whereas far away from the cores the energy
density gradually drops down to zero in an expanding universe. This
corresponds to the delay of end of inflation $\delta N \gg 1$, and to
density perturbations $\delta\rho/\rho> 1$.  If these topological
defects were isolated objects immersed in empty Minkowski space,
one can show that their  thickness is much {\it smaller} than their
Schwarzschild radius~\cite{LL}. Therefore one would come to
the conclusion that there is no suppression for the probability of
black hole formation in this scenario. However, it turns out that
this is not what happens.

To investigate this issue note that the sinusoidal wave (\ref{top1})
is a particular example of a perturbation $\delta\psi$ which is
symmetric with respect to $\psi \leftrightarrow  - \psi$. The specific
shape of these perturbations indicates that one cannot study them
by considering Gaussian distributions centered away from
$\psi = 0$. However, one could apply
the theory of density perturbations to those parts of the defects
away from $\psi = 0$. The main ingredient is the calculation of the
time delay for rolling of the field $\psi$ down to the minimum of its
effective potential. For the centers of topological defects this time
delay is infinite, which corresponds to density perturbations
$\delta\rho/\rho \gg 1$. Thus, one may expect important black hole
formation near the centers of  inflating topological defects.
However, as we will see, the mere fact that topological defects
correspond to large perturbations does not automatically imply
that they form black holes, because the distribution of density
near the cores of these defects is extremely non-Gaussian.
Note that topological defects correspond to places where the
distribution of the field $\psi$ near their centers grows linearly,
$\psi(x) \sim H^2x$, see Eq.~(\ref{top1}). Therefore the density
distribution near the topological defects looks like a narrow
peak. Meanwhile, the distribution of the scalar field produced by
the usual density perturbations at $\psi \not = 0$ has
relatively flat regions where the spatial derivative of the field
$\psi$ vanishes. These places have greater size, for the same
amplitude of density contrast, and thus can be more easily
converted into black holes.

To estimate the density contrast $\delta\rho/\rho$ in the vicinity
of the topological defects we will use the fact that during the last stages
of inflation the field $\phi$ in our model is already close to the
minimum of the effective potential. Thus, let us consider the field
(\ref{top3}) at the moment when it reaches the minimum of its
effective potential at  $\psi = f$.  This happens approximately at a
time $t_e$ given by equation $(H/\sqrt 2\pi)\, e^{m_\psi t_e} = f$.
At this moment the distribution of the field $\psi$, as a function of
the distance $r = x e^{-Ht}$ from the center of the topological defect,
can be written in a very crude approximation as
\begin{equation}\label{ top4}
\delta\psi_c \sim { H\over \sqrt 2\pi}\,\exp{(m_\psi\, t_e) }\,\sin(H r) \ .
\end{equation}
Let us rewrite its evolution in terms of $f$, counting time from $t_e$.
For small $r$,
\begin{equation}\label{ top5}
\delta\psi_c(r) \sim  f H r\ \exp{(m_\psi\, t) }\ ,
\end{equation}
which remains a good approximation until $\delta\psi_c(r)\sim f$.
The time delay until the field at a distance $r$ reaches $f$ is
\begin{equation}\label{top5}
\delta N(r) = H \delta t(r)  = -\,{H\over m_\psi}\, \ln Hr \ .
\end{equation}
This means that the density of the topological defect will exceed
the average density by $\delta\rho/\rho \sim \delta N(r) \sim 1$
only in the core  of the topological defect, at a distance
\begin{equation}\label{top6}
r \sim H^{-1} \exp\Bigl(-{m_\psi\over H}  \Bigr) \ .
\end{equation}
For the usual inflating topological defects considered in Ref.~\cite{LL}
one has $m_\psi/H \ll 1$, and therefore the deviation of density
from the average density becomes large for the whole region from
$r \sim H^{-1}$ to $r = 0$. That is why one may expect such
topological defects to look like huge black holes from the outside.
Meanwhile in the model of  Ref.~\cite{Guth} one has $m_\psi/H >1$.
As a result, the part of the volume inside the horizon where the
energy density deviates considerably from that on the horizon will
be extremely small, being suppressed by $\exp(-3m_\psi/H)$.
Therefore the total mass excess on the horizon scale is too small to
cause the black hole formation. The same behavior of the density
distribution is repeated at smaller scales as well. This
suggests that inflating topological defects in the theories with
$m_\psi > H$ do not necessarily lead to dangerous black hole formation.

The validity of this argument depends on many assumptions. For
example, if the universe is matter dominated then black holes are
formed even if the density perturbations are not very large. Therefore
topological defects may serve as seeds for small black holes which
are formed at soon after inflation, before the universe reheats and
becomes radiation dominated. This may lead to very interesting
consequences, to be discussed in Sect.~6. Meanwhile, black holes
which are formed in the models with a prolonged second stage of
inflation are very large, and they are  formed very late, when the
universe is supposed to be radiation dominated. Our investigation
indicates that large black holes do not   come from topological defects.
However,  they may be produced by the usual density perturbations,
which will be studied in the next section.

We should emphasize that our conclusions concerning black hole
formation from inflating topological defects should be considered
only as very preliminary. Indeed, here we deal with perturbations
of density which are extremely large, $\delta\rho/\rho \gg 1$, and
our simple analysis using methods of perturbation theory may be
inapplicable. It is sufficient to remember  that the interior of topological
defects in our model continues to inflate for a long time after inflation
ceased to exist in the outer space. Thus we have here a complicated
wormhole-type geometry which requires a very careful analysis,
similar to the numerical investigation performed in Ref.~\cite{Sakai}.
It is the first time when we need to know a detailed structure of the
universe filled by a gas of inflationary topological defects on the
scale smaller than the size of the observable part of the universe.
We need to return to investigation of this fascinating question in the
future. In the meantime we can only say that it would be premature
to conclude that the model of Ref. \cite{Guth} contradicts
observational data  solely on the basis of our investigation of
density perturbations produced by inflating topological defects.

For this reason in the next section we will concentrate on  the theory
of usual (non-topological) density perturbations in models with two
scalar fields and apply it to our model. The results that we have
obtained from the investigation of topological defects suggest that,
in the model under consideration, usual density perturbations may
also happen to be very large.    In what follows we will show that
this is indeed the case.

\section{Density perturbations}

Quantum fluctuations of the scalar fields are responsible for
curvature perturbations on a comoving hypersurface at the end of
inflation, which can be evaluated as the change in the time (or number
of $e$-folds) it takes to end inflation. In the case of adiabatic
perturbations, e.g.  in single-field inflation driven by the field
$\psi$, the amplitude of the curvature perturbation (on equal energy
density hypersurfaces) remains fixed on super-horizon scales, so it
can be calculated as $\delta N = [H\delta\psi/\dot\psi]_*$ at horizon
crossing where $\delta\psi$ can be estimated as $H/2\pi$
\cite{GuthPi,Mukhanov}. The origin of this curvature perturbation is
that the jump of the field in the direction opposite to its motion
leads to a time delay in the end of inflation which can be estimated
by $\delta t = \delta\psi/\dot\psi$.  This equation immediately
suggests that in the hybrid inflation model where the second stage of
inflation after the phase transition begins at $\psi = 0$, density
perturbations on the length scale corresponding to the moment of the
phase transition should be extremely large, since at that time
$\dot\psi = 0$.  This is the standard situation in all models where
inflation occurs near the local maximum of the effective potential. It
does not lead to any troubles if the corresponding scale is many
orders of magnitude greater than the present size of the observable
part of the universe.  But in the model of Ref.~\cite{Guth} this scale
was supposed to be rather small, $l_c < 10$ Mpc.  One certainly does
not want to have the observable part of the universe densely populated
by huge black holes.

However, these considerations are too naive.  First of all, in our
case we have two fields moving, $\phi$ and $\psi$, so even if the
field $\psi$ does not move at all, the whole field configuration
evolves in time. Therefore the delay of the end of inflation is no
longer given by the simple one-field expression $\delta t =
\delta\psi/\dot\psi$. Also, in the presence of more than one field,
the perturbations may not be adiabatic and, as a consequence, their
amplitude need not be constant~\cite{Wands,deltaN2}. The effect of the
perturbation must be integrated along the perturbed trajectory to
the end of inflation, or until the evolution becomes
adiabatic~\cite{deltaN1,deltaN2}.

Finally, because we no longer have a unique trajectory in field
space, we must consider which trajectories are likely to be realized
in practice. Because of the quantum fluctuations, the field $\psi$
does not stay exactly at the point $\psi = 0$ at the moment of the
phase transition and we must investigate the dispersion of the
probability distribution to see what the likely amplitude of density
perturbations will be.

\subsection{Single-field perturbations}

Firstly we consider the simplest regime, before the phase transition,
where single-field results apply.
At large values of $\phi\gg\phi_c$ the $\psi$ field has a large positive
mass and remains fixed at $\psi=0$. The amplitude of $\psi$
fluctuations crossing outside the horizon is negligible. Thus we need
only consider adiabatic fluctuations $\delta\phi_*$ along the
trajectory, given by Eq.~(\ref{deltaphi}), which perturb the time it
takes to end inflation,
\begin{equation}\label{dNphi}
\delta N = \left[ {H\delta\phi\over\dot\phi} \right]_*
 = {C(r) \, H\over2\pi \,r\,\phi_*} \, .
\end{equation}
The power spectrum of curvature perturbations on comoving
hypersurfaces, ${\cal R} = \delta N$, is then given by
\begin{equation}\label{PR}
{\cal R}^2(N) = {C(r)^2\over4\pi^2}\,{\gamma\over\beta r^2}\,
e^{2r(N-N_c)}\,,
\end{equation}
where $N_c$ is the number of $e$-folds from the phase transition
($\phi=\phi_c$) to the end of inflation (which we have seen could
be of order 20--30).
Assuming these perturbations are responsible for the observed
temperature anisotropies in the microwave background, they give a
constraint on the parameters of the model.  The small (of order 1TeV)
Hubble constant during inflation implies that the contribution of
gravitational waves to the microwave background anisotropies will be
negligible~\cite{LL93,CLLSW}. The low multipoles
of the angular power spectrum measured by COBE~\cite{COBE} give a
value ${\cal R}^2 \simeq 3\times 10^{-9}$ on the scale of our current
horizon, corresponding to
\begin{equation}
N_{\rm CMB} \simeq 46 +{2\over3}\,\ln\left( {M\over10^{11}{\rm GeV}}
\right) + {1\over3}\,\ln\left({T_{\rm rh}\over10^7{\rm GeV}}\right)\,.
\end{equation}
For parameters $\alpha$ and $\beta$ or order one, this
requires $\gamma \lesssim~10^{-8}$. Note that for these parameters we
will have $\phi_c \geq 10^4\, {\rm TeV}\gg H_0$.

This is one of the few cases in inflationary cosmology where we have an
almost exact expression for the amplitude of curvature
perturbations~\cite{short}.
The only approximation we have made is to assume that the energy density
remains constant, $\dot\phi^2/2+m^2\phi^2/2 \ll M^4$, so that we
can neglect the backreaction on the metric.  Using Eq.~(\ref{dNphi}),
we see that this will be true as long as
\begin{equation}
{8C(r)^2\over3r\, {\cal R}^2}\,{M^4\over M_{\rm P}^4} \ll 1\, .
\end{equation}
It follows that the allowed range of $r$ is
\begin{equation}
{8\over3{\cal R}^2}\,{M^4\over M_{\rm P}^4}
 \ll r \ll {3\over2} - \sqrt{8\over9\pi\,{\cal R}^2}
  \,{M^2\over M_{\rm P}^2}\,,
\end{equation}
which for $M\ll 10^{16}$GeV essentially leaves $\alpha$ as a free
parameter in the range $0<\alpha<9/4$. At the upper limit,
$r\sim3/2$, the correction coefficient $C(r)$ becomes large, giving
a significant amplification of the curvature perturbations ${\cal R}$
compared to the usual slow-roll approximation.

Because the comoving scale at horizon crossing is just proportional
to the scale factor $\propto e^N$, the scale dependence of the power
spectrum in Eq.~(\ref{PR}) readily gives the tilt of the spectrum as
\begin{equation}
n-1 \equiv {d\ln{\cal R}^2\over d\ln k} = 2r \, ,
\end{equation}
which becomes $2\alpha/3$ in the slow-roll limit. Note that in
principle for this model one could have any value of the tilt in
the range $1<n<4$. A precise measurement of $n$, which may be
possible with the next generation of satellite
experiments~\cite{PSI}, would give a tight constraint on $\alpha$.
Present limits give $n=1.2\pm0.3$~\cite{COBE}. Since $n$
could be greater than one, there are also limits coming from
production of black holes at small scales, $n<1.4$~\cite{Carr}.
Together they give $\alpha\leq0.6$, which constrains the size of
the correction coefficient in Eq.~(\ref{PR}) to lie in the range
$0.9\leq C<1$.

\subsection{Two-field perturbations}

In this subsection we describe the interesting regime in which
the system goes through the phase transition and quantum fluctuations
of both fields become important. Here the curvature perturbation on
a comoving hypersurface at the end of inflation
cannot be given simply in terms of that at horizon crossing since it does
not remain constant on super-horizon scales. In order to compute the
amplitude of the curvature perturbation at the end of inflation, one usually
integrated the coupled differential equations for the two fields'
fluctuations and evaluated their amplitude at the end of inflation from
that  at horizon crossing. Only recently, Sasaki and Stewart~\cite{deltaN1}
developed a formal method for computing the metric perturbations at
a given hypersurface from the change in the number of $e$-folds to that
hypersurface as a {\em local} function in field space, in the slow-roll
approximation.
This method was shown in Ref.~\cite{deltaN2} to be equivalent to the
usual method of integrating the quantum field fluctuations. Unless one
finds solutions for all trajectories in field space, the problem
remains analytically intractable. In Ref.~\cite{deltaN2} we found
particular cases, with separable potentials for the interacting fields,
in which the field trajectories were integrable and we could
write explicit expressions for the amplitude of curvature perturbations
at the end of inflation. In principle, for a general model all we need is a
computer to evaluate the change in the number of $e$-folds to the
end of inflation due to quantum fluctuations of the fields, for all points
in field space. This gives us the possibility to investigate density
perturbations even for very complicated theories where it is not possible to
express the final result  in a compact  analytical form. In the
case of hybrid inflation we are fortunate to have a complete analytical
solution for the classical evolution in Regions~I and~II which takes us
from $\phi=\phi_c$ to the end of inflation. This will allow us to
compute in a compact way the amplitude of curvature perturbations
at the end of inflation (where the system becomes adiabatic and
the amplitude remains constant on super-horizon scales) in terms of
the field fluctuations at horizon crossing.

The perturbation in the number of $e$-folds from any point
$(\phi_*,\psi_*)$ in Region~I to the end of inflation can be computed
by evaluating the number of $e$-folds from a given point up until a
surface of
constant $\psi=\psi_m$ in Region~II, since the time from $\psi_m$ to
the end of inflation is fixed, see Eq.~(\ref{psiN}). Due to the
overlap between the regions, this surface will also lie in Region~I
for $\psi_m\ll\phi_c=\sqrt{\beta/\gamma}H_0$. The number of $e$-folds
to $\psi_m$ is given by
\begin{equation}\label{Nphim}
N(\phi_*,\phi_m) = N_m - \, {1\over r}
\ln\left({\phi_*\over\phi_m}\right)\, ,
\end{equation}
where $\phi_m$ is a function of the trajectory parametrized by $c_1$,
\begin{equation}
\phi_m \simeq (c_1 \, A)^{r/s}\, \psi_m^{-r/s} \, .
\end{equation}
Using the solution for $\psi(\phi)$ in Region~I, given by
Eq.~(\ref{growpsi}), we can eliminate $c_1$ to give
\begin{equation}\label{Nphipsi}
N(\phi_*,\psi_*) \simeq {1\over s}\ln\left[{\psi_e\over\psi_*}\right]
+ \, {1\over s}\ln\left[{\phi_*^\nu\,Y_\nu(\kappa\phi_*)
\over A}\right]\, ,
\end{equation}
only in terms of the fields at horizon crossing. This is one of the few
cases in which such an integration can be done completely up to the
end of inflation, see also Ref.~\cite{deltaN2}.

Note that in general a trajectory beginning at a perturbed point
$(\phi+\delta \phi,\psi +\delta\psi)$ may end up at a completely
different point in field space compared with the nonperturbed
trajectory. This can make the comparison of the lengths of the
trajectories very complicated and could lead to entropy as well as
curvature perturbations at the end of inflation. However, in our case
all trajectories merge at the end of inflation, and this complication
does not arise. The fact that by the end of inflation we are left with
a single field (and thus all perturbations have become adiabatic)
allows us to equate the amplitude of curvature perturbations on a
comoving hypersurface at the end of inflation with perturbations on
comoving hypersurfaces at late times and, in particular, at the
surface of last scattering, see Ref.~\cite{deltaN2}.

We can now evaluate the perturbation in the number of $e$-folds as
\begin{equation}\label{dN}
\delta N \simeq {\kappa\over s}\, {Y_{\nu-1}
(\kappa\phi_*)\over Y_\nu(\kappa\phi_*)} \, \delta\phi_*
+ {\delta\psi_*\over s\psi_*} \,.
\end{equation}
Note that $\,Y_\nu(z) \sim z^{-\nu}\,$ for small $z$, and thus the first
term vanishes if the point $(\phi_*,\psi_*)$ lies in Region~II, which
gives $\delta N = \delta\psi_*/s\psi_*$, as required by Eq.~(\ref{psiN}).

For $\phi_*\simeq\phi_c$ we have $\partial N/\partial\phi_* \simeq
1/\phi_c \ll \partial N/\partial\psi_*$, for $\psi_*\ll\phi_c$. Since the
quantum fluctuations $\delta\phi_*$ and $\delta\psi_*$ at this time
are both of order $H/2\pi$, the amplitude of curvature perturbations
is given by
\begin{equation}\label{dNphic}
{\cal R} = \delta N \simeq {C_k\,H \over 2\pi \,s\, \psi_*} \,,
\end{equation}
where $C_k\simeq1$, see Eq.~(\ref{Ck}).
On the other hand, for
$\phi_*\ll\phi_c$, the coefficient $C_k\gg1$, see Fig.~1, which is
then responsible for large curvature perturbations.

It is clear from Eq.~(\ref{dNphic}) that for small values of $\psi$,
the amplitude of curvature perturbations can become arbitrarily
large. In Fig.~2 we show a few equal-$N$ surfaces  in field space
$(\psi/H_0,\phi/\phi_c)$, around and below $\phi=\phi_c$, together
with the line $\delta N = 1$.

The amplitude of density perturbations on a comoving hypersurface
when the curvature perturbations (\ref{dNphic}) re-enter the
horizon is given by
\begin{equation}\label{DP}
{\delta\rho\over\rho} = {2+2w\over5+3w}\,{\cal R} \,,
\end{equation}
where $p=w\rho$ is the equation of state of the universe at re-entry.
We therefore expect large density perturbations on scales
associated with the phase transition.

\subsection{Quantum diffusion}

So far we have discussed the classical evolution of the homogeneous
field and the effect of perturbations about the classical
trajectories on a given scale for values of $\phi$ above $\phi_c$ and
below $\phi_c$. However at some stage the role of quantum diffusion
of the coarse-grained field $\psi$ on super-horizon scales dominates
over its classical motion. Purely classical trajectories in Region~I
beginning with $\psi\simeq0$ above $\phi_c$ are focussed along
$\psi=0$, due to the large effective mass of $\psi$ at large $\phi$,
and continue to evolve close to $\psi=0$ long after the point
$\phi_c$ when it becomes an unstable ridge. In practice we require
quantum diffusion of the $\psi$ field to move the field off the ridge
and begin its roll down to the global minimum.

If diffusion washes out any trace of the classical motion as we cross
$\phi=\phi_c$, it does not make sense to calculate the curvature
perturbations in terms of the classical trajectories. This would
destroy our notion of associating points in field space with a given
number of $e$-folds from the end of inflation. Quantum diffusion close
to $\phi_c$ could distort equal time hypersurfaces so much that we
lose information about the origin of trajectories. This problem is
analogous to that of trying to trace the path of photons beyond the
(cosmological) last scattering surface. Beyond this surface, photons
scatter many times and an observer at late times can no longer
associate their pathlengths with a single smooth surface.  Thermal
diffusion is responsible for this loss of information in the
trajectories of photons beyond last scattering. In our case it is
quantum diffusion of the scalar field that determines the loss of information.
Note that in the above mentioned sense the region near the critical point
$\phi = \phi_c$ becomes opaque to wavelengths equal to the wavelengths of
perturbations formed at $\phi = \phi_c$, but it remains transparent to
perturbations with much greater and much smaller wavelengths.

As argued above,
our calculation of density perturbations relies only on being able
to associate a given scale at late times, determined by the number of
$e$-folds $N$, with a unique smooth surface in field-space.
Our results for the amplitude of perturbations near $\phi=\phi_c$ in
Eqs.~(\ref{dN}) and~(\ref{dNphic}) show that there is indeed a region
near $\phi = \phi_c$, $\psi = 0$ for which $\delta N$ becomes very
large. In Fig.~2 the region under the $\delta N=1$ curve is the
dangerous region. The question now is whether the parameters of the
model are such that this affects a significant number of trajectories
in field space. To determine this we should evaluate the probability
distribution for the scalar fields and calculate how much of the
initial wave-packet suffers large perturbations in the number of
$e$-folds, i.e. $\delta N\geq 1$.  A heuristic constraint could be
that the size of the scattering region should be much smaller than the
size of the wave-packet (very much like scattering of light of
wavelength $\lambda$ by a target with a diameter $a< \lambda$).  These
$\delta N\sim1$ perturbations correspond to large curvature
perturbations on asymptotic comoving hypersurfaces, which later become
black holes. If a significant part of the packet enters the region
where $\delta N>1$, then at late times we cannot reconstruct the
amplitude of the initial perturbation, corresponding to large scales
($\phi>\phi_c$). What happens is that, due to quantum diffusion,
different scales will mix and their amplitudes will be undetermined
for an asymptotic observer at late times.

According to \cite{self}, in single field slow-roll inflation  the regime of
$\delta N \geq~1$
can be identified with quantum diffusion dominating over classical
motion, $\delta\phi \geq \dot\phi/H$, i.e. with the well known
self-reproduction regime~\cite{Vilenkin,self}. However, in two-field
inflation this may no longer be the case. For example, fluctuations
in one of the fields may not affect the time taken to end inflation.
Even in our Region~II, where only $\psi$ determines $N$, it is the
asymptotic time delay that determines $\delta N$, not the instantaneous
perturbation $[\delta\psi/\dot\psi]_*$ at horizon crossing.

Let us now calculate the probability distribution  for the field $\psi$ in
Region~I, as the field $\phi=\phi_c\,e^{-r(N-N_c)}$ slowly rolls down
its potential. This can be done using the stochastic approach to
inflation~\cite{Stochastic,Guth}. Assuming a initial delta distribution for
$\psi$ at $\phi\gg \phi_c$, and an average quantum diffusion per Hubble
volume per Hubble time $\approx H/2\pi$,\footnote
{Note that our earlier analysis shows that this is an over-estimate
for $\phi\geq\phi_c$. However it should give a safe upper bound on the
dispersion of the wave-packet.}
the time-dependent probability distribution has the form
\begin{equation}
P(\psi,t)={1\over\sqrt{2\pi}\,\sigma}\,e^{-\psi^2/2\sigma^2(t)}\,,
\end{equation}
where the dispersion $\sigma^2(t)$ satisfies the evolution equation
\begin{equation}
{d\sigma^2(t)\over dt} = {H^3\over4\pi^2} + {2\beta H\over3}
\left(1-{\phi^2\over\phi_c^2}\right) \sigma^2(t)\,.
\end{equation}
Under a change of variables, $x\equiv\exp[-2r(N-N_c)]$ and $\,S(x) \equiv
\sigma^2(t)/H^2$, this equation becomes
\begin{equation}
{d S\over dx} = -{1\over8\pi^2rx} - {\beta(1-x)\over3rx}\,S(x)\,,
\end{equation}
which has an exact solution
\begin{equation}\label{dispersion}
S(x) = {1\over8\pi^2r}\, \left({e^x\over ax}\right)^a\Gamma(a,ax)\,,
\end{equation}
where $a\equiv\beta/3r\simeq\beta/\alpha$ is a constant and
$\Gamma(a,z)$ is the Incomplete Gamma function. The solution
$S(N)$ characterizes the dispersion of the classical trajectories due
to quantum fluctuations. Since the region where $\delta N\simeq 1$
has a width $\psi\simeq H/2\pi s$, see Eq.~(\ref{dNphic}), at $\phi=
\phi_c$, most classical trajectories will pass through this region.
It is still possible that just one $e$-fold after the phase transition the
distribution will have spread so much that only a small fraction of
all the trajectories still goes through this region. We thus consider
the dispersion of the $\psi$-distribution one $e$-fold after the phase
transition, when $x=\exp(-2r)$. Note that the $\delta N \geq1$
region is broader there, since $C_{k=e}>1$ from Fig.~1, but the
distribution (\ref{dispersion}) is also wider. If the spread of the
probability distribution is still within the $\delta N \geq1$ region at
this stage, i.e.
\begin{equation}\label{diffcond}
\langle\psi^2\rangle
 = {H^2\over8\pi^2r}\, \left[{\exp(e^{-2r})\over a\,e^{-2r}}\right]^a\,
\Gamma(a,a\,e^{-2r}) < {H^2\over 4\pi^2 s^2} \,,
\end{equation}
it will be difficult to avoid large perturbations on this scale. Note that
this condition is totally independent of the coupling $\gamma$.
We have plotted this condition in Fig.~3, where we show the
contour plots of the dispersion of the distribution in units of the
size of the $\delta N=1$ region, $\langle\psi^2\rangle^{1/2} =
n\,H/2\pi s$, for $n=1,\dots,7$ in parameter space $(\alpha,\beta)$.
The figure shows is that in order for the distribution to have spread
(one e-fold {\em after} the phase transition) several times the size of
the $\delta N = 1$ region one needs $\beta \gg 1$. The large
separation of the lines indicates how difficult it is for the distribution to
spread, unless $\beta \gg 1$. If they were closely packed it would mean
that, for not very large $\beta$, the distribution would be much wider
than this dangerous region and it would be possible for most trajectories
to avoid this region. As it stands, for the values of the parameters
in Ref.~\cite{Guth}, most of the trajectories will go through this region.

In summary, we have shown that for $\alpha<0.6$ (the range of parameters
allowed by observations of the spectral tilt on large and small scales)
we find large perturbations, $\delta N\sim1$, along most trajectories
at the phase transition, unless $\beta\gg1$. This corresponds to large
curvature perturbations on these scales and thus to large density
perturbations after inflation which, as we shall see,
leads inevitably to the formation of black holes.

\section{Black hole production}

\subsection{Probability of black hole formation}

We have seen that quantum fluctuations of the fields can be
responsible for large curvature perturbations on a comoving
hypersurface at the end of inflation. These perturbations, on scales
that left the horizon 20 -- 30 $e$-folds before the end of inflation,
as in the model we study,
re-enter the horizon during the radiation era and could in principle
collapse to form primordial black holes.
The theory of production of primordial black holes from initial
inhomogeneities was first discussed in Ref.~\cite{HCarr}, see
also~\cite{Khlopov}. There is an expression for the probability that a
region of mass $m$, with initial density contrast $\delta(m)\equiv
\delta\rho/\rho|_m$, becomes a primordial black hole,
\begin{equation}\label{PBH}
P(m) \sim \delta(m)\,e^{-\bar\beta^4/2\delta^2}\ ,
\end{equation}
where $\bar\beta^2 \sim w$~\cite{HCarr}. In the derivation of this
equation it was assumed that the universe was a barotropic fluid
$(p=w\rho$) during gravitational collapse, and that the initial
density contrast $\delta(m)$ is much smaller than one. There seems to
be disagreement over the value of the parameter $\bar\beta$ and the
way to calculate it in a radiation dominated universe, see
Ref.~\cite{Guth}. Novikov et al.~\cite{Novikov,NovFrol} give a simpler
prescription based on numerical calculations. They claim that even for
density perturbations less than one at horizon crossing, a black hole
will form as long as the perturbation in the metric $\delta g_{ab} =
2\Phi \,\delta_{ab}$ is of order $0.75 - 0.90$, where $\Phi$ is the
gauge-invariant Newtonian potential~\cite{NovFrol}. The analysis of a
probability distribution for density perturbations with a peaked
spectrum is beyond the scope of this paper, but we expect a
probability distribution like that of Eq.~(\ref{PBH}) with
$\bar\beta^2 \lesssim 1$, somewhat larger than that of
Ref.~\cite{HCarr}. Furthermore, Carr points out in Ref.~\cite{HCarr}
that, for a scale invariant spectrum with density contrast
$\,\delta\sim1$ at horizon crossing, the probability of black hole
formation is $P \sim 1/2$ on all scales, and half of the mass of the
universe is always in the form of primordial black holes.  From the
above discussion it is clear that for the large density contrasts
produced during the phase transition, $\delta =(4/9)\,\delta N \sim
0.5$ at horizon crossing, see Eq.~(\ref{dNphic}), there is no
suppression of the probability of black hole formation on scales
associated with the phase transition.

Note that in order to calculate the precise probability of black hole
formation (\ref{PBH}), we have to solve the ambiguity in the value of
the parameter $\bar\beta$.  This would require a much more detailed
investigation, see e.g. \cite{NovNas}.  However, for our purposes it
is enough to realize that in a Gaussian distribution where the typical
fluctuations of the density contrast are about $0.5$, the fluctuations
$\delta \sim 1$ are just one standard deviations away from $\delta
\sim 0.5$. Therefore for every $10$ horizon-sized regions with density
contrast $\delta=0.5$ we typically find one region with $\delta
\sim1$, which can be expected to collapse and form a black hole.

We should note that the previous discussion of the probability of the
black hole formation is based on investigation of conventional
inflationary density perturbations. In our case, in addition to such
perturbations, we have a dense gas of inflating topological defects. One
might expect each of them to become a black hole, which would make the
number of black holes much greater even than the one suggested by the
estimates based on Eq.~(\ref{PBH}). (Here we are talking about the
monopole-like inflating topological defects, since domain walls and
strings with symmetry breaking $f \sim M_{\rm P}$ lead to a
cosmological disaster even if they do not inflate and form
black holes.)  However, we believe that this issue requires a more
detailed analysis, see Sect. 4, and therefore in this paper we impose
on the model of Ref.~\cite{Guth} only those constraints which follow
from our investigation of the usual density perturbations.

Let us now evaluate the typical size and mass of the black holes
produced by these perturbations. Suppose that after the phase
transition the universe inflated $e^{N_c}$ times.  Then at the end of
inflation the physical scale that left the horizon during the phase
transition is $H_0^{-1} e^{N_c}$, where $H_0$, as before, is the Hubble
constant during inflation. Suppose that soon after inflation the equation
of state became $p = \rho/3$, as for the ultrarelativistic gas. Then the
scale factor of the universe after inflation grows as $\sqrt{tH_0}$.
The scale $H_0^{-1} e^{N_c} \sqrt{tH_0}$ becomes comparable to the
particle horizon $\sim t$ at
\begin{equation}\label{FORMATION}
t_h = H_0^{-1} e^{2N_c} \,
\end{equation}
when the energy
density becomes smaller than the inflationary energy density $\sim
H_0^2M_{\rm P}^2$ by a factor $e^{-4N_c}$. At that time perturbations with
density contrast $\delta \sim 1$ form black holes of size $H_0^{-1}
e^{2N_c}$ and mass
\begin{equation}\label{BHMASS}
M_{\rm BH} \simeq {M_{\rm P}^2\over H_0}\,e^{2N_c} \,.
\end{equation}
For $N_c \sim 30$ one would have black holes with mass $\sim
10^{37}$~g, comparable to the masses of the black holes in the centers
of galaxies. This is a very interesting mass scale, but copious
production of such black holes would lead to catastrophic cosmological
consequences.

By changing the parameters of our model one can make the duration of the
second stage of inflation rather short. For small black holes with $N_c = {\cal
O}(1)$ in Eq.~(\ref{FORMATION}) should be
modified because they are formed at the stage when the energy may
still be dominated by the oscillations of the inflaton field with the
equation of state $p = 0$. This changes a little our estimate for the
time of formation of the black hole,
\begin{equation}\label{FORMATIONsmall}
t_h = H_0^{-1} e^{3N_c} \,
\end{equation}
and for the black hole mass,
\begin{equation}\label{BHMASSsmall}
M_{\rm BH} \simeq {M_{\rm P}^2\over H_0}\,e^{3N_c} \,.
\end{equation}
The smallest black holes would correspond to $N_c \sim 1$ (and $H_0
\sim 10^3$~GeV, as in \cite{Guth}) would have a mass of about
$10^{11}$~g. Perturbations $\delta \sim 1$ giving rise to black holes
in the mass interval $10^{11} -10^{37}$~g are clearly ruled out by the
bounds of Ref.~\cite{Carr}. Thus, we should avoid at all costs the
dangerous region $\delta N \simeq 1$, since otherwise we will have too
many large black holes.

A very interesting possibility arises when one considers such
a peak in the spectrum, for not very massive black holes. From the
bounds of Ref.~\cite{Carr} we see that density contrasts of order
$\delta\sim 3\bar\beta/20$ are just enough to give $\Omega_0=1$
in the mass range $10^{15} - 10^{30}$~g.
Taking $\bar\beta\simeq 1$ from Ref.~\cite{NovFrol} and using
Eq.~(\ref{delta}), we find that a parameter $s=3$ could indeed
give the desired density contrast. This corresponds to a
bare mass for the triggering field, $m_\psi \simeq 4$~TeV, which is
very natural.  Furthermore, the associated mass scale can be
computed from
$N_c = 32/3 \sim 11$ as $M_{\rm BH} = 2 \times 10^{20}~{\rm g} =
10^{-13}~M_\odot$. Using the average density of our galaxy,
$\rho_g \sim 10^{-25}~{\rm g/cm}^3$, we find that these small black holes
may populate the halo of our galaxy and be separated from
each other an average of $10^{15}$~cm or about 6 times the size of the
solar system. They could very well constitute the missing mass in
our galaxy, and still pass undetected by the microlensing
surveys~\cite{MACHO}.

Note  that changing slightly the parameters
of the model one changes simultaneously the scale and the height of the
peak in the black hole spectrum. This means that numerical values of the black
hole masses and the distances between them can be made substantially different
by modification of the hybrid inflation model. These numbers are very
sensitive to the details of the theory of black hole formation, which still
requires a more complete analysis.  It is important, however, that in the
context of the hybrid inflation scenario the possibility that black holes may
contribute to the dark matter of the universe becomes quite realistic.

The idea that dark matter may consist of black holes produced after
inflation was explored earlier by Ivanov, Naselsky and
Novikov~\cite{NovNas}.  They performed a detailed investigation
of the probability of formation of large black holes in such models, and in
this respect their work can be extremely useful. Their model required the
existence of a plateau
in the effective potential, which would lead to a high peak in the
spectrum of density perturbations.  However, it is very difficult
to obtain a realistic model of a single scalar field where one has almost
exactly flat spectrum
$\delta\rho/\rho \sim  5\times10^{-5}$ on all scales from $10^{28}$ to
$10^{20}$ cm, and a sharp peak with $\delta\rho/\rho \sim 1$ on a
slightly smaller scale.   Meanwhile, as we have shown, in the hybrid inflation
scenario this possibility emerges in a very natural way.

\subsection{Reheating from black hole evaporation}

A very interesting application of the above results comes when we
consider a two-stage inflation with a sufficiently short period of inflation
after the phase transition. With the parameters of the model~\cite{Guth}
even the smallest black holes are very heavy and evaporate too late.
However, by choosing a model of hybrid inflation with a sufficiently large
Hubble constant and short second stage of inflation one can have a very
interesting regime when the black holes will dominate the energy density of
the universe soon after the end of inflation and later evaporate before
nucleosynthesis, reheating the universe.

Let us first assume that small black holes were formed in a radiation dominated
universe, soon after the usual stage of reheating after inflation.
To evaluate the probability of black hole formation  we need to know  the
density contrast at horizon crossing during
the radiation dominated stage, see Eqs.~(\ref{dNphic}) and~(\ref{DP}),
\begin{equation}\label{delta}
\delta = {4\over 9}\,\delta N \simeq {2C_k\,H_0\over 9\pi s\,\psi} \simeq
{4\over 9 s}\,.
\end{equation}
The number of $e$-folds in the second stage of inflation is given by
$N_c = (1/s)\ln(\psi_e/\psi)$. With initial condition  $\psi \sim H_0/2\pi$
we have
\begin{equation}
e^{N_c} \sim \left({2 M_{\rm P}\over s H_0}\right)^{1/s}\,.
\end{equation}
The time it takes a black hole to evaporate  is given by
\begin{equation}
\tau \simeq {1\over g^*M_{\rm P}} \left({M_{\rm BH}\over M_{\rm P}}\right)^3
\, .
\end{equation}
Here $g^* \sim 10^2$ is the effective number of particle species at
the time of the black hole evaporation.  For black holes with small
$N_c$ formed at the matter dominated stage we have
\begin{equation}\label{EVAP}
\tau \sim   {1\over g^*M_{\rm P}} \left({M_{\rm P}\over H_0}e^{3N_c}
\right)^3\, .
\end{equation}

Suppose that the fraction of matter in the black holes initially was
only very small, and the universe was radiation dominated
from the time of black hole formation to the time they evaporate. Then
the fraction of mass in black holes grows during this time as $a(t)
\sim \sqrt t$ due of the more rapid decrease of the energy density of
relativistic particles outside black holes. At the instant before the
black holes finally evaporate, the fraction of energy in black holes
has grown by a factor of
\begin{equation}\label{growth}
\sqrt{\tau\over t_h} \sim { M_{\rm P}\over \sqrt{g^*}  H_0} e^{N_c/2} \,.
\end{equation}
Therefore even if only a small fraction of energy was in the black
holes initially, because the probability of their formation was
suppressed by the exponential factor in Eq.~(\ref{PBH}), we only require
$P(\delta)>(\tau/t_h)^{-1/2}$ for the black holes to come to dominate
the energy density of the universe before they evaporate.

To give a particular example, let us consider hybrid inflation with
$H_0 \sim 10^{14}$ GeV (which is much greater than in the model of
Ref. \cite{Guth}). Let us take, e.g., $s \sim 3$, i.e. larger than the
usual parameters of the models of Ref.~\cite{Guth} but much smaller
than those of Ref.~\cite{hybrid}. Then we have the total number of
e-folds at the second stage of inflation $e^{N_c}\approx 40$. The
density contrast in this case is $\delta \sim 1/6$. The fraction of
matter in the black holes, according to (\ref{PBH}), will be about
$10^{-5}$.  In fact, this is a rather conservative estimate, since our
investigation of topological defects suggests that this number may be much
greater. The black holes will be produced at the moment
$t_h \sim 6\times 10^{-34}$ s. They will have mass $M_{\rm BH} \sim
6\times 10^4$~g, and will evaporate at $\tau \sim 2\times 10^{-16}$~s,
much earlier than the epoch of nucleosynthesis. Because of the large
growth of the scale factor and redshift of energy of relativistic
particles, at the time of the black hole evaporation practically all
matter in the universe will be in black holes.  This means that
practically all particles which exist in the universe at $t >
10^{-16}$~s are created at the moment of the black hole evaporation.

Note that in the above example, although the probability
of black hole formation is very small, they still give the dominant
contribution to the energy density at late times because the energy of
relativistic particles decreases much faster than that of black holes.
However, this black hole dominance may begin much earlier if they are
formed before conventional reheating is complete and the equation of state
is $p \simeq 0$.  This condition can be easily satisfied in the case of
very small black holes. Then there will be no exponential suppression
of the probability of the black hole production (\ref{PBH}), and the
fraction of energy in the black holes could be large from the very
beginning.

The process of black hole evaporation could be responsible for the
baryon asymmetry in the universe, even though it is not very easy to
get large baryon asymmetry by this mechanism~\cite{Liddle}.
Typically it is assumed that reheating and thermalization of the universe
occurs due to the inflaton field decay and the subsequent particle
interactions, or through bubble collisions like in first order inflation.
The natural assumption was that the gravitational interaction
at the stage of reheating could be neglected. Here we have another,
very unusual mechanism of reheating. Even in the absence of bubble
wall collisions or a large coupling of the inflaton to matter, a
considerable fraction of matter after inflation could be in the form of
small black holes. Unlike in the extended inflation scenario~\cite{Hsu},
in our case all such black holes are formed in the same mass range
given by Eqs.~(\ref{BHMASS}) and~(\ref{BHMASSsmall}).
If the probability of black hole formation is not strongly suppressed,
then very soon they dominate the energy density of the
universe~\cite{Liddle}. Eventually the evaporation of these black
holes could reheat the universe.
This opens up an interesting possibility of connecting
the origin of matter in the universe with black hole physics.

Let us estimate the reheating temperature of the universe in this
scenario.  Black hole masses in the process of their evaporation
decrease as $M_{BH}\bigl[1-t/\tau\bigr]^{1/3}$. (Here we have taken
into account that the age of the universe $t_h$ at the moment of the
black hole formation is much smaller than their evaporation time
$\tau$.)  The main part of the energy release by the evaporating black
holes occurs at the end of the time interval $\tau$.  Therefore one
may simply use the standard temperature-time relation for the hot
universe to get the following estimate of the reheating temperature
$T_r$ after the black hole evaporation:
\begin{equation}\label{reheat}
T_r^2 \sim {M_{\rm P} \over 4\pi \tau} \sqrt{45\over g^*\pi} = {\sqrt {45 g^*}
H^3\over 4\pi\sqrt \pi M_{\rm P} } e^{-3 N_c} \ .
\end{equation}
For a particular example which we studied ($H \sim 10^{14}$ GeV, $s
\sim 3$, $g^*\sim 10^2$) we get $T_r \sim 2\times 10^9$~GeV.  This
estimate is extremely sensitive to the choice of the parameters. One
can easily get reheating temperature as high as $10^{10}$~GeV or even
greater, or as small as $1$~eV.  The only real constraint on this
temperature is that one should be able to produce the baryon asymmetry
of the universe during or after black hole evaporation, and before
nucleosynthesis. This picture differs considerably from the standard
theory of reheating due to the decay of the inflaton field, see
e.g.~\cite{reh1,reh2}.

Perhaps one can appreciate a potential importance of this regime if one
remembers that the standard reheating due to the inflaton decay often is very
inefficient because of the small coupling of the inflaton to matter
\cite{reh1,book}. In such cases the universe for a long time remains matter
dominated ($ p = 0$).  In some other cases reheating is extremely efficient in
the
very beginning, but later becomes inefficient, so that the universe
eventually may become  matter dominated again \cite{reh2}. But then
formation of black holes is no longer suppressed by radiation pressure.
In this case matter easily collapses
into small black holes, which later evaporate and reheat the universe. Thus,
the absence of the usual reheating triggers black hole formation, which
eventually leads
to a very efficient reheating of the universe. This is a win-win situation,
where
black holes can reheat the universe even if the standard reheating
mechanism is inoperative!

\section{``Natural'' Hybrid Inflation}

As we have seen, the model (\ref{hybrid}), as well as the version
proposed in \cite{Guth}, lead to a copious formation of huge black
holes if one requires that (unlike in the original version of hybrid
inflation) there is an additional inflationary stage after the phase
transition. This problem occurs because typical classical trajectories
in this model go very close to $\psi = 0$. One can avoid this problem
by a modification of the shape of the effective potential
\cite{Lazarides}. Also, as we have shown above, black hole production
can be even useful if the second inflationary stage is very short and
the black holes are very small. But there exists another problem,
which we will consider now together with the first one.

The main reason why many authors are trying to implement hybrid
inflation in supersymmetric theories is to protect the flatness of the
effective potential in the $\phi$-direction. One may try to relate the
small mass of the field $\phi$ to the gravitino mass $m_{3/2} \sim 1
$~TeV, which appears because of supersymmetry breaking. If one argues
that the parameter $M^2$ in Eq.~(\ref{supernatural}) is of the order
of the intermediate scale of supersymmetry breaking $m_{3/2} M_{\rm P}$,
then there appears to be no unexplained small parameters in the
model. Still the appearance of the term $M^4 \sim (m_{3/2} M_{\rm P})^2
\cos^2(\psi/\sqrt2 f)$ in (\ref{supernatural}) remains somewhat
unclear to us.  If one expands $M^4\cos^2(\psi/\sqrt2 f)$ in powers of
$\psi$ for $M \sim 10^{11}$~GeV and $f \sim 10^{18}$~GeV as in
Ref.~\cite{Guth}, one would find an extremely small coupling constant
$M^4/f^4 \sim 10^{-30}$ in front of the term $\psi^4$.  It was pointed
out in \cite{Guth} that such couplings may appear in a natural way if
one introduces certain superpotentials which lead to nonrenormalizable
interactions. However, to study nonrenormalizable terms in an
internally consistent way it would be necessary to consider models
based on supergravity, which was outside the scope of our
investigation, as well as of the investigation performed
in~\cite{Guth}.

Fortunately, both smallness of the parameter $M^4$ and the shape of
the potential can be explained if one interprets $\psi$ as a pseudo
Goldstone field similar to the axion field. One may consider a model
of a complex scalar field $\Psi(x) \equiv (f(x)/\sqrt 2)\, \exp i
\theta(x)$,~which after spontaneous symmetry breaking can be
represented as $(f/\sqrt 2)\, \exp(i \psi(x)/ f )$. If the original
effective potential was a function of $\Psi^*\Psi$, the field $\psi$
will be massless. However, nonperturbative (instanton or wormhole)
effects may give this field a small mass (see Ref.~\cite{KLLS} for a
recent discussion of this issue). This effect can be described by
adding operators breaking initial $U(1)$ symmetry of the effective
potential. Consider the family of operators $g_n
(\Psi \pm \Psi^*)^n f^{4-n}$. Since these operators appear because of
nonperturbative effects, the coupling constants $g_n$ may be exponentially
small. One may take, for example, the simplest operator $g_1 (\Psi \pm
\Psi^*) f^{3}$, and add to it a constant term $\sqrt 2 g_1 f^4$
normalizing the vacuum energy to zero. This gives the effective
potential of the field $\psi$, which is completely analogous to the
standard axion potential,\footnote
{Note that in~\cite{CLLSW} a model of hybrid inflation was studied
based on a supersymmetric Peccei-Quinn model. In that case it was
the self-interaction energy of the radial degree of freedom $f(x)$ that
gave the constant potential energy in the false vacuum.}
\begin{equation}\label{naturalpot}
V(\psi) =  2\sqrt 2 g_1 f^4 \cos^2\left({\psi\over  2f}\right) \ .
\end{equation}
Note that in this potential $\psi/f$ is an angular variable
from $0$ to $2\pi$.  This potential coincides with the effective
potential of the field $\psi$ in Eq.~(\ref{supernatural}) up to an
obvious change $2\sqrt 2 g_1 f^4 \to M^4$, $f \to f \sqrt 2$. (Our
definition of $f$ corresponds to a canonical normalization of the
field $\phi$ kinetic terms.) In this context both the shape of the
potential for the field $\psi$ and the smallness of the term $M^4
\cos^2\left({\psi/\sqrt2 f}\right)$ are explained in a natural
way. Potentials of this type have been used in ``natural inflation''
models \cite{BG,Natural}.  The problem with ``natural inflation'' is
that for the ``natural'' value of symmetry breaking $f {\
\lower-1.2pt\vbox{\hbox{\rlap{$<$}\lower5pt\vbox{\hbox{$\sim$}}}}\ }
M_{\rm P}$ inflation is too short and the spectrum index $n$ is
significantly less than $1$.  There is no such problem in our
model; the main purpose of the introduction of the field $\psi$ is to
support inflation {\it before} the phase transition rather than after
it.

Thus it makes a lot of sense to explore cosine potentials such as
Eq.~(\ref{naturalpot}) in the context of hybrid inflation. But with
the interpretation of the field $\psi$ as a pseudo Goldstone particle,
one cannot couple it to the field $\phi$ in the way proposed in
Eq.~(\ref{supernatural}).  Now $\psi$ is the angular part of the field
$\Psi$, and one cannot write any superpotentials for the fields $\Psi$
and $\phi$ which would result in the simple interaction terms $\sim
\psi^2\phi^2$. However, since we already reinterpreted the cosine term
in Eq.~(\ref{supernatural}) as appearing from the anomalous term $g_1
(\Psi +\Psi^*) f^{3}$, we can go further and introduce an anomalous
interaction term $g_2 (\Psi e^{-i\theta} -\Psi^*e^{i\theta})^2 f^{2}
\phi^2$. Note that we have introduced here for generality the phase
shift $\theta$ between the two anomalous terms.  Indeed, the origin of
the first and of the second anomalous terms may be different, and {\it
a priori}\, one should not expect $U(1)$ symmetry to be broken by
these two terms in the same way. In what follows we will assume that
$\theta$ is small. The resulting effective potential including the
mass term of the field $\phi$ can be represented in the following
form:
\begin{eqnarray}\label{doublesupernatural2}
V(\phi,\psi) &=&
2 {\lambda_1^2}  f^4 \cos^2{\psi\over 2f}
+ {\lambda_2^2\over 2}f^2 \phi^2\sin^2\Bigl({\psi\over f} - \theta\Bigr)
\nonumber
\\
&+& {m^2\over2} \phi^2  \, .
\end{eqnarray}

Let us now analyze the shape of this potential and its relation to the
more usual hybrid inflation potential (\ref{hybrid}). Consider first
the case $\theta = 0$. At large $\phi$ the dominant term involving
$\psi$ in Eq.~(\ref{doublesupernatural2}) is the second one, which
implies that at large $\phi$ the field $\psi$ will settle in one
of the minima at $\psi/f=n\pi$, where $n$ is some integer. For odd
values of $n$ the energy density due to the self-interaction
cosine-squared term also vanishes and we are left with conventional
chaotic inflation with $V=m^2\phi^2/2$. However for even values of
$n$, the self-interaction term is non-zero and $\psi$ is trapped in a
false vacuum, like the model in Eq.~(\ref{hybrid}). Near $\psi = 0$
the potential in Eq.~(\ref{doublesupernatural2}) is given by
\begin{equation}\label{doublesupernatural3}
V(\phi,\psi)  =
2 {\lambda_1^2 }   f^4 -  {\lambda_1^2 f^2\over 2}\psi^2     +
{\lambda_2^2\over 2
} \phi^2 \psi^2
+ {m^2\over2} \phi^2  \, .
\end{equation}
One concludes, that about $\psi=0$ the bare mass squared of the field
$\psi$ is $m_\psi^2 = -\lambda_1^2 f^2$, but that the effective
mass-squared becomes positive for $\phi > \phi_c = \lambda_1
f/\lambda_2$.  Near $\phi_c$ at $\psi = 0$ the energy density is given
by $2\lambda_1^2\,f^2 ( f^2+ m^2/4\lambda_2^2)$. The first term
dominates, as in the usual hybrid inflation, for $m \ll \lambda_2 f$,
and we have exactly the same as the model we have analyzed in the
preceding sections where
\begin{equation}\label{lastHubble}
H = \sqrt  {\pi\over 3}\  {4\lambda_1  f^2  \over M_{\rm P}} \ ,
\end{equation}
and the dimensionless parameters introduced in Section~\ref{DYN}
are given by $\alpha=(3/16\pi) m^2M_{\rm P}^2 / \lambda_1^2f^4$,
$\beta=(3/16\pi) M_{\rm P}^2/f^2$ and $\gamma=\lambda_2^2$.  The
curvature perturbations produced at $\phi>\phi_c$ are then given
by~Eq.(\ref{PR}). For small $\alpha$ we have~\cite{hybrid}
\begin{equation}\label{hybrid2}
{\lambda_2 \lambda_1^2 f^5\over M_{\rm P}^3 m^2} \simeq
2 \times 10^{-6}\, .
\end{equation}
in order to agree with the COBE normalization~\cite{COBE}.

There are two regimes one may consider in this theory. First of all,
one may assume that the coupling constants are not extraordinarily
small. Then, as was shown in Ref.~\cite{hybrid}, the conditions $m \ll
H$ and $\delta\rho/ \rho \sim 10^{-5}$ imply that there was no second
stage of inflation in this model, i.e. everything is going on as in
the first version of the hybrid inflation scenario~\cite{hybrid}. One
may consider, for example, the following parameters:
$f \sim 10^{16}$~GeV (GUT scale),
$m = 10^{10}$~GeV (intermediate SUSY breaking scale),
and $\lambda_1 \sim \lambda_2 \sim 10^{-3}$.
In this case all conditions mentioned above will be satisfied. The
parameters $\alpha \sim 10^{-2}$ and $\beta \sim 10^5$ so there
will be no second stage of inflation and no anomalous black hole
production. For $\theta = 0$ there will be domain wall
production. However, the domain walls formed in this scenario are
unstable because they are always bounded by strings with $\Psi = 0$.
Moreover, it is sufficient to consider models with a very small
non-zero value of $\theta$, so that all the evolution will go in one
direction, and there will be no domain walls or other topological
defects. One can easily understand this if one takes into account that
at large $\phi$ the minimum of the effective potential with respect to
the field $\psi$ is at $\psi = \theta f \not = 0$.  Thus it is enough
to have $\theta > 10^{-15}$ to avoid black hole formation in our
model.

As one might expect, the same model for a different choice of
parameters can reproduce all the results of the model proposed in
\cite{Guth}, including the second stage of inflation after the phase
transition. This requires to take $\lambda_1^2 \sim 10^{-30}$, which
is an extremely small number. However, in our case the existence of this
small parameter is not surprising, because it could appear due to
nonperturbative effects. Typically the value of this parameter is
suppressed by factors such as $\exp{(-8\pi^2 /g^2)}$ where $g$ is the
gauge coupling constant. In some models this suppression may not be
very significant, but in general this suppression can easily give
numbers much smaller than $10^{-30}$ \cite{KLLS}.  In particular,
in the usual axion theory with $f \sim 10^{12}$ GeV the corresponding
constant is of the order of $10^{-130}$. {}From this perspective it is
more surprising that in this model the coupling constant $\lambda_2^2$
is not required to be equally small. This disparity can be easily
alleviated if one does not insist that the masses of both fields as well
as the Hubble constant at the end of inflation should be of the order
of $m_{3/2}$.

For $\theta = 0$ in this model, just like in the model of
Ref.~\cite{Guth}, one obtains inflating topological defects, very
large density perturbations on the scale corresponding to the moment
of the phase transition, and catastrophic black hole
production. However, it is no longer a generic property of the
model. Remember that the dangerous area of the phase space is located
very close to $\phi = \phi_c$ and $\psi = 0$. For example, large
density perturbations are generated only at $\psi < H \sim 10^{-15}
f$, for $H \sim 10^3$ GeV and $f \sim 10^{18}$ GeV.  It is enough to
have $\theta > 10^{-15}$ to avoid black hole formation in our model.
Thus for generic values of $\theta$ inflationary trajectories never
come close to $\psi = 0$, and the problem disappears.
For non-integer values of $\theta/\pi$ hybrid inflation can occur
along $\psi=\theta+n\pi$ for any integer value of $n$, with the false
vacuum energy density equal to $2\lambda_1^2f^4\cos^2(\theta/2)$ for
even $n$ or $2\lambda_1^2f^4\cos^2(\theta/2)$ for odd values.

We do not want to pretend that the ``natural'' hybrid inflation model is
necessarily very natural. In order to study this question one would
have to investigate the appearance of the anomalous terms in a more
detailed way, and to analyse the possible effects of adding a more
standards terms like $\Psi^*\Psi\phi^2$. Our main purpose was to show
that the cosine terms with small coefficients can be incorporated into
the hybrid inflation models, and that it is possible to avoid the
problem of large density perturbations and black hole production
in these models. However, as we
have argued in the previous section, primordial black holes produced
after inflation under certain conditions may lead to very interesting
cosmological consequences. On the other hand, there is a simpler way
to get rid of the black holes; one may simply return to the original
hybrid inflation scenario without the second inflationary stage.

\section{Conclusions}

Inflationary theory was first proposed about 15 years ago, and its
main principles are by now well understood. It is therefore
surprising to see that slight modifications in simple and natural
models may lead to important and sometimes absolutely unexpected
consequences. For almost fifteen years we knew that inflation
exponentially dilutes the density of topological defects, but only two
years ago did we learn that topological defects may inflate themselves
\cite{LL}.  It was also thought that old inflation did not work, and
chaotic inflation always predicted that the universe is flat. A year
ago, however, it was found that the simplest hybrid model where one
takes old inflation potential for one of the fields and chaotic
inflation potential for another one (even if these two fields do not
interact with each other!) leads to the universe consisting of
infinitely many separate universes with all possible values of $\Omega
\leq 1$ \cite{Open}. Now we encountered one more surprising fact. For
many years it seemed clear that inflation erased all pre-existing
inhomogeneities and did not leave much room for the production of
primordial black holes, which had been the subject of active
investigation in the end of the 70's. Now we see that in a very simple
inflationary model one can easily obtain a large amount of black
holes.  They are formed only in a specific mass range, determined by
the duration of inflation after the phase transition.  Typically they
are huge, but depending on the parameters of the model they can be
very small as well.

Note that black holes are not necessarily a curse but could also be a
blessing. In the simplest model studied here the probability
of black hole formation is not suppressed at all, and their number
appears to be too large.  We propose some modifications of this model
where black hole formation is strongly suppressed.  It is possible
for certain values of the parameters of the model to have the
right amount of relatively
large black holes, $M\sim 10^{15} - 10^{30}$~g, that have not yet
evaporated and may be responsible for the dark matter in the halos
of galaxies. In a particular model considered in Sect. 6 we have
shown that the halo of our galaxy may consist of black holes of mass
$\sim 10^{21}$ g. However,  numerical  values of the masses and
abundances of the black holes are strongly model dependent.
Depending on the parameters of the models there may be just enough black holes to have $\Omega_0=1$ in the
universe. If the black holes are supermassive, one could  speculate about
their relation to the black holes in the centers of galaxies.

Suppressing the number of large black holes down to a desirable
level requires a certain degree of fine tuning. But it is relatively easy
to make the black holes very small and harmless by making  the
second stage of inflation short and by ending inflation at large $H$.
For $H \sim 10^3$ GeV the smallest black hole masses $\sim
{M_{\rm P}^2/H}$ are still very large, about $10^{11}$ g, and they
evaporate very late, at $t \sim 10^3$ s . But if, e.g.,  one takes the
models with $H \sim 10^{14}$ GeV, one can obtain black holes
with a   mass $\sim 6\times 10^4$ g, evaporating at $t \sim 10^{-16}$ s.
Such black holes would  dominate the universe after their formation
until evaporation. Evaporation of black holes may lead to baryon
asymmetry production. During the last fifteen years this
mechanism of baryogenesis was largely ignored since it seemed
impossible to produce many small black holes after inflation (see
however Ref.~\cite{Liddle}). We may now return to the investigation
of this interesting possibility. Independently of the issue of
baryogenesis, one should emphasize that the possibility of the
black hole dominance at the intermediate post-inflationary stage may
change completely the mechanism of reheating after inflation, which
would proceed via black hole evaporation.

In the course of our work we have further developed a method of
investigation of density perturbations which can be applied even for
complicated systems of several coupled scalar
fields~\cite{deltaN1,deltaN2}. This method is rather simple and
powerful. It gives analytical results in those cases in which the
motion in field space is integrable, like in hybrid inflation and in
theories with coupled inflaton and dilaton fields, but it can be used
in a more general context as well. The method consists of three main
parts.  First of all, for any point in the $(\phi, \psi)$ space one
finds (either analytically or numerically) an inflationary trajectory
going from this point, and calculates the number of e-folds
$N(\phi,\psi)$ for this trajectory. This problem is easy to solve
numerically even for very complicated potentials. Then one perturbs
the position of the initial point $(\phi,\psi)$ by adding to it
inflationary jumps, which typically are of the order $H/2\pi$, but may
be greater or smaller, see Section~3. This gives us the perturbation
of the number of e-folds $\delta N$, which is directly related to the
density perturbations: $\delta\rho/\rho = (4/9)\delta N(\phi,\psi)$ at
re-entry during the radiation dominated era. Note that the resulting
density perturbations for a given $N$ (i.e. for a given wavelength)
will depend on the place $(\phi,\psi)$ our trajectory came from. Thus
the remaining step is to evaluate the probability that for a given
number of e-folds $N$ from the end of inflation the field was at any
particular point $(\phi,\psi)$. This problem can be solved by using
the stochastic approach to inflation \cite{Stochastic}. This approach
tells us what is the probability to find density perturbations of a
given amplitude with a given wavelength.\footnote{Note that a similar
problem of evaluation of probability appears also in the theory of a
single inflaton field if one takes into account the possibility of
nonperturbative effects due to large jumps along the inflationary
trajectory \cite{infloids}.  However, in the theory of several scalar
fields this issue becomes of more immediate importance.} Usually at
the last stages of inflation we have a single inflationary trajectory,
independent of initial conditions. In our case, however, inflationary
trajectory is unstable at $\phi < \phi_c$, $\psi = 0$; all
trajectories bifurcate due to quantum fluctuations. Therefore the
calculation of the probability distribution was necessary to show that a
{\it typical} amplitude of density perturbations produced near the
point of the phase transition is very large.

As an important by-product of our investigation we have found a new
type of inflating topological defect. They appear in the models where
the curvature of the effective potential is somewhat greater than
$H^2$, but nevertheless the time necessary for symmetry breaking to
occur is much greater than $H^{-1}$. These defects do not inflate
eternally and do not form a fractal structure found
in~\cite{LL}. Still inflation in the cores of these defects continues
for a while even after it ends outside of them.  As a result, they
lead to large density perturbations of a specific type.  Until now
inflating topological defects could be considered as an interesting
but somewhat esoteric feature of certain inflationary
models. Typically the distance from us to these defects was many
orders of magnitude greater than the size of the observable part of
the universe. They were important for understanding of the global
structure of the universe, but not of our local
neighborhood. However, in hybrid inflation models with
two stages of inflation these defects are abundantly produced at the
moment of the phase transition, and populate the part of the universe
which is accessible to our observations.  We believe that the new type
of inflating topological defects deserves separate investigation. It
would be very interesting to understand whether they lead to
black hole formation and to explore other possible observational
consequences of these exotic objects. It is amazing that very simple
models of two scalar fields can exhibit such a rich and interesting
behavior!

\section*{Acknowledgements}
The authors thank Ed Copeland and Andrew Liddle for
collaboration at an early stage of the work.
One of us (A.L.) is very grateful to Lev Kofman
for persistent encouragement of investigation of post-inflationary
primordial black holes and to Slava Mukhanov for  useful comments.
We are especially grateful to Alan Guth and Lisa Randall for
clarifying discussions of the model of  Ref.~\cite{Guth}.
J.G.B. and D.W. thank the Physics Department
at Stanford University, where this work was done, for their wonderful
hospitality.  The work of A.L. was supported by NSF grant PHY-9219345.
J.G.B. and D.W. acknowledge support from PPARC (U.K.).  This work was
supported in part by a NATO Collaborative Research Grant,
Ref.~CRG.950760.

\

\

\begin{center}
{\bf FIGURE CAPTIONS}
\end{center}

\noindent
{\bf Figure 1:} The amplitude of quantum fluctuations of the symmetry
breaking field $\delta\psi$ in units of $H/2\pi$, as a function of scale $k$,
evaluated when this scale left the horizon.  Here $k_c$ corresponds to
the scale that left the horizon during the phase transition, when $\phi=
\phi_c$. The figure corresponds to generic values of the parameters,
($\alpha=.3,\beta=8$). The asymptotic value as $k\to\infty$ is
$C_\infty=8.91$.

\vspace*{24pt}
\noindent
{\bf Figure 2:}  The thick dashed-dotted line corresponds to the
$\delta N = 1$ region within which it is not possible to define equal-time
hypersurfaces, and density perturbations are of order one. We also
show a few equal-number-of-$e$-folds contours in field space
($\psi/H,\phi/\phi_c$), for generic values of the
parameters, ($\alpha=.3,\beta=8$). The dashed sections enter the
$\delta N = 1$ region where due to quantum fluctuations we cannot
associate a definite number of $e$-folds to the end of inflation.

\vspace*{24pt}
\noindent
{\bf Figure 3:}  Contour lines for the dispersion of the $\psi$ distribution
at the phase transition (dashed line) and one $e$-fold after the phase
transition (continuous line), compared with the size of the
$\delta N=1$ region, $\langle \psi^2 \rangle^{1/2} = n\,H/2\pi s$
(with $n=1, \dots, 7$ from the bottom up) in parameter space
($\alpha,\beta$). In the region below the curves, the probability
distribution for the $\psi$ field cannot avoid the dangerous $\delta N =1$
region and large-amplitude perturbations will be expected at scales
associated with the phase transition.

\end{document}